\documentclass{aastex631}
\let\tablenum\relax
\usepackage{siunitx}
\usepackage{gensymb}
\usepackage[T1]{fontenc}
\DeclareUnicodeCharacter{2212}{-}
\newcommand\chandra{{\it Chandra}}
\newcommand\xmm{{\it XMM-Newton}}
\newcommand\nustar{{\it NuSTAR}}

\newcommand\swift{{\it Swift}}

\newcommand\xrism{{\it XRISM}}

\newcommand\FBQS{FBQS~J163302.6+234928}
\newcommand\nineteenfivetwoeight{2MASXi~J0729087+400836}

\newcommand\PG{PG~1302-102}
\newcommand{\uat}[2]{\href{http://astrothesaurus.org/uat/#2}{#1 (#2)}}

\begin{document}

\title{
NuSTAR Observations of Candidate Subparsec Binary Supermassive Black Holes
}

\correspondingauthor{M. Lynne Saade}
\email{mlsaade@usra.edu}

\author[0000-0001-7163-7015
]{M. Lynne Saade}
\affiliation{Department of Physics and Astronomy, University of California, 475 Portola Plaza, Los Angeles, CA 90095}
\affiliation{Science \& Technology Institute, Universities Space Research Association, 320 Sparkman Drive, Huntsville, AL 35805, USA}

\author[0000-0002-8147-2602]{Murray Brightman}
\affiliation{Cahill Center for Astronomy and Astrophysics, California Institute of Technology, 1216 E. California Blvd., Pasadena, CA 91125}

\author[0000-0003-2686-9241]{Daniel Stern}
\affiliation{Jet Propulsion Laboratory, California Institute of Technology, 4800 Oak Grove Drive, Pasadena,
CA 91109}

\author[0000-0002-7898-7664]{Thomas Connor}
\affiliation{Harvard-Smithsonian Center for Astrophysics, 60 Garden Street, Cambridge, MA 02138}

\author[0000-0002-0603-3087]{S.G. Djorgovski}
\affiliation{Cahill Center for Astronomy and Astrophysics, California Institute of Technology, 1216 E. California Blvd., Pasadena, CA 91125}

\author[0000-0002-1271-6247]{Daniel J. D'Orazio}
\affiliation{Niels Bohr International Academy, Niels Bohr Institute, Blegdamsvej 17, 2100 Copenhagen, Denmark}

\author[0000-0002-5956-851X]{K.E.S. Ford}
\affiliation{Department of Astrophysics, American Museum of Natural History, Central Park West at 79th Street, New York, NY 10024}
\affiliation{Department of Science, Borough of Manhattan Community College, City University of New York, New York, NY 10007}
\affiliation{Physics Program, The Graduate Center, City University of New York, New York, NY 10016}

\author[0000-0002-3168-0139]{Matthew J. Graham}
\affiliation{Cahill Center for Astronomy and Astrophysics, California Institute of Technology, 1216 E. California Blvd., Pasadena, CA 91125}

\author[0000-0003-3633-5403]{Zolt\'an Haiman}
\affiliation{Department of Astronomy, Columbia University, New York, NY 10027}

\author[0000-0003-1470-5901]{Hyunsung D. Jun}
\affiliation{SNU Astronomy Research Center, Astronomy Program, Dept. of Physics and Astronomy, Seoul National University, Seoul 08826, Republic of Korea}'

\author[0000-0002-0273-218X]{Elias Kammoun}
\affiliation{IRAP, Universit\'{e} de Toulouse, CNRS, UPS, CNES 9, Avenue du Colonel Roche, BP 44346, F-31028, Toulouse Cedex 4, France}
\affiliation{INAF – Osservatorio Astrofisico di Arcetri, Largo Enrico Fermi 5, I-50125 Firenze, Italy}

\author[0000-0002-0765-0511]{Ralph P. Kraft}
\affiliation{Harvard-Smithsonian Center for Astrophysics, 60 Garden Street, Cambridge, MA 02138}

\author[0000-0002-9726-0508]{Barry McKernan}
\affiliation{Department of Astrophysics, American Museum of Natural History, Central Park West at 79th Street, New York, NY 10024}
\affiliation{Department of Science, Borough of Manhattan Community College, City University of New York, New York, NY 10007}
\affiliation{Physics Program, The Graduate Center, City University of New York, New York, NY 10016}

\author[0000-0001-8121-0234]{Alexei Vikhlinin}
\affiliation{Harvard-Smithsonian Center for Astrophysics, 60 Garden Street, Cambridge, MA 02138}

\author[0000-0001-5819-3552]{Dominic J. Walton}
\affiliation{Centre for Astrophysics Research, University of Hertfordshire, College Lane, Hatfield AL10 9AB, UK}
\affiliation{Institute of Astronomy, University of Cambridge, Madingley Road, Cambridge CB3 0HA, UK}

\begin{abstract}
We present analysis of \nustar{} X-ray observations of three AGN that were identified as candidate subparsec binary supermassive black hole (SMBH) systems in the Catalina Real-Time Transient Survey based on apparent periodicity in their optical light curves. Simulations predict that close-separation accreting SMBH binaries will have different X-ray spectra than single accreting SMBHs. We previously observed these AGN with \chandra{} and found no differences between their low energy X-ray properties and the larger AGN population. However some models predict differences to be more prominent at energies higher than probed by \chandra{}. We find that even at the higher energies probed by \nustar{},  the spectra of these AGN are indistinguishable from the larger AGN population. This could rule out models predicting large differences in the X-ray spectra in the \nustar{} bands. Alternatively, it might mean that these three AGN are not binary SMBHs.
\end{abstract}

\keywords{\uat{Seyfert Galaxies}{1447};
\uat{X-ray active galactic nuclei}{16};
\uat{Quasars}{1319};
\uat{Supermassive black holes}{1663}}

\section{Introduction} \label{sec:intro}

Binary supermassive black holes (SMBHs) are expected to be a ubiquitous consequence of galaxy mergers. When two galaxies merge, their corresponding SMBHs will pair up into binaries. The binary separation will shrink due to gravitational interactions with stars \citep{2006ApJ...642L..21B,2017MNRAS.464.2301G} and gas \citep{2007Sci...316.1874M} in the merged galaxy. When the binary reaches sub-parsec separation, gravitational waves become the dominant mode by which the binary shrinks, allowing the two black holes to spiral together and merge \citep{1980Natur.287..307B}. In the process they will release gravitational waves that could be detected by future observatories such as the {\it Laser Interferometric Space Antenna} \citep[{\it LISA};][]{2017arXiv170200786A,2023LRR....26....2A}, as well as pulsar timing arrays \citep{2021ApJ...915...97X}. 

Candidate binary SMBHs have been identified in active galactic nuclei (AGN) through a variety of methods, including unusual jet morphologies \citep[e.g.,][]{2005A&A...431..831L,2013MNRAS.428..280C,2013ApJ...779...41T,2014MNRAS.445.1370K,2019MNRAS.482..240K}, emission line profiles \citep[e.g.,][]{2012ApJS..201...23E,2013ApJ...777...44J,2013ApJ...775...49S,2014ApJ...789..140L,2016ApJ...822....4L,2019MNRAS.482.3288G}, candidate periodic features in AGN light curves \cite[e.g.,][]{1996ApJ...460..207L,2015Natur.518...74G,2015MNRAS.453.1562G,2016MNRAS.463.2145C,2019ApJ...884...36L,2022ApJ...926L..35O,2020MNRAS.499.2245C,2021MNRAS.500.4025L,2022arXiv220611497C}, and X-ray variability \citep[e.g.,][]{2020ApJ...896..122L,2020ApJ...902...10S}.  For a recent review of this field, see \citet{2022LRR....25....3B}. Of the candidates identified, the current strongest candidate is OJ 287, which displays periodic flares that are well explained by a model in which a secondary black hole passes through a primary black hole's accretion disk once per decade \citep{1996ApJ...460..207L}.  This model has been used to predict a flare in OJ 287 down to the precision of a day \citep{2008Natur.452..851V}. A flare was observed in 2020 that is consistent with the binary model \citep{2020MNRAS.498L..35K}. However a predicted flare in 2022 was not seen, and the disk luminosity was found to be a factor of 10-100 times lower, indicating some aspects of the original model need modification, such as including precession and/or considering a lower mass for the primary SMBH \citep{2023MNRAS.522L..84K}. 

The second strongest candidate is PG 1302-102, which shows some evidence for consistent periodicity in its optical \citep{2015Natur.518...74G}, ultraviolet \citep{2015Natur.525..351D,2020MNRAS.496.1683X}, and mid-infrared light curves \citep{2015ApJ...814L..12J}, as well as in the precession of its radio jet \citep{2018A&A...615A.123Q}. In particular, the ratio of the amplitudes in the UV and optical matches expectations under the assumption that the sinusoidal variation is due to relativistic Doppler modulation from binary orbital motion given the UV and optical spectral slopes \citep{2015Natur.525..351D, 2020MNRAS.496.1683X}.

Another potential way to detect binary SMBHs is through X-ray emissions. Since X-rays probe the portion of an AGN closest to the central black hole(s), the presence of a subparsec SMBH binary could potentially have a large imprint on the high-energy emission. Several  models have predicted the X-ray emissions of binary SMBH AGN, though they differ in their predictions in part because current simulations are unable to simulate thin disks and thus make ad-hoc approximations about the thermodynamics. Some analyses predict a notch in the X-ray spectrum \citep[e.g.][]{2018MNRAS.476.2249T} and/or for the spectral shape of the high-energy continuum to be harder \citep[e.g.,][]{2014ApJ...785..115R,2017ApJ...835..199R,2018MNRAS.476.2249T,2019ApJ...879..110K}, while others predict more modest differences, if any \citep[e.g.,][]{2018ApJ...865..140D,2022ApJ...928..137G}. Many models predict increased X-ray luminosity as well \citep[e.g., 10x higher in the 10-100 keV band,][]{2015MNRAS.447L..80F}.

In our previous paper, \citet{2020ApJ...900..148S} (henceforth referred to as SA20), we observed 7 AGN identified as potentially periodic by \citet{2015MNRAS.453.1562G} from the Catalina Real-Time Transient Survey \citep[CRTS;][]{2009ApJ...696..870D}.  We used \chandra{} observations to test theoretical models of accreting binary SMBHs and potentially determine whether the AGN were binary SMBHs. We did not find any significant differences between the spectra of these AGN and the spectra of single-SMBH AGN. While there are many possible reasons for this result (discussed at length in SA20), one potential reason is that the differences in X-rays could be modest in the soft X-rays, but more dramatic in the harder X-rays, as predicted by some calculations \citep[e.g.][]{2014ApJ...785..115R}.

Three of the AGN in \citet{2020ApJ...900..148S} have \nustar{} \citep{2013ApJ...770..103H} observations through a combination of our own proposal and archival data. In this work we analyze these \nustar{} spectra to see if there is any evidence that the predicted differences between binary SMBH AGN and single SMBH AGN show up in the hard X-rays. The structure of our paper is as follows: in Section 2 we describe the X-ray data and reduction thereof. In Section 3 we discuss the X-ray properties of our sample and compare it to other samples in the literature. In Section 4 we discuss the implications of our results in the context of both theory and observations. For calculating the luminosities, we use the cosmology used in NASA's Extragalactic Database (NED), namely $\Omega_{\rm M} = 0.308$, $\Omega_\Lambda = 0.692$ and $H_0 = 67.8\, {\rm\,km\,s^{-1}\,Mpc^{-1}}$.

\section{Observations and Data Analysis}
The X-ray observations used in this paper are listed in Table \ref{tab:obstable}. We use all available \nustar{} data for the objects in the SA20 sample. This amounted to three objects: \nineteenfivetwoeight{}, \PG{}, and \FBQS{}. \nineteenfivetwoeight{} and \PG{} have \nustar{} data from our Cycle 6 proposal (P.I. M. Saade). The \FBQS{} \nustar{} observations were obtained from the archive (P.I. E. Kammoun).

For soft X-ray data, we preferentially used data  taken simultaneously  with  \nustar{}. For \PG{}, this was \swift{} \citep{2004ApJ...611.1005G} data; for \FBQS{}, this was \xmm{} \citep{2001A&A...365L...1J} data. \nineteenfivetwoeight{} did not have any simultaneous observations, so we re-use the \chandra{}  \citep{2002PASP..114....1W} observation reported in SA20. All the high-energy observations were background-subtracted  and fit in XSPEC \citep[][version 12.12.1]{1996ASPC..101...17A}. The spectra were grouped by a minimum of one count per bin. In this situation, XSPEC uses a modified version of the C-statistic known as the W-statistic. Below we describe the specific details of each observatory's data analysis.

\subsection{\nustar{}}
We reduced and extracted the \nustar{} data with HEASOFT (version 6.30.1), NuSTARDAS (version 2.1.2), and \nustar{} CALDB (version 20220525). We used 40\arcsec\ radius circular regions centered on each source for the extraction, and 100\arcsec\ radius background regions. In fitting the spectra {of \nineteenfivetwoeight{} and \PG{},} we fixed the cross-normalization constant of FPMA to 1.0, and that of FPMB to 1.04, where the latter is based on calibration observations of the bright source 3C~273 reported in \citet{2015ApJS..220....8M}. We also did this for the first \nustar{} observation of \FBQS{}. For the second observation of \FBQS{}, we let the FPMA constant freely vary, and fixed the FPMB constant to be $1.04\times$ the FPMA constant. The \nustar{} background dominates the source above 30 keV, so we used the 3-30 keV range for the spectral fitting.

\subsection{\swift{}}
The HEASARC archive includes a \swift{} observation contemporaneous with the \nustar{} observation of \PG{}. We reduced and extracted the XRT data with HEASOFT (version 6.30.1), the \swift{} XRT CALDB (version 20210915), and the \swift{} XRT Data Analysis Software (version 3.7.0). We used a circular source region of 25\arcsec{} radius and a circular background region of 50\arcsec{} radius. We used the 0.3-10 keV range for the spectral fitting.

\subsection{\xmm{}}
\FBQS{} has a simultaneous \xmm{} observation for both \nustar{} observations. We extracted the data using the \xmm{} Scientific Analysis Software (version 20.0.0). For all \xmm{} cameras we used circular source regions 20\arcsec{} in radius. We used a circular background region 80\arcsec{} in radius for the MOS cameras and 60\arcsec{} in radius for the PN camera. The latter was smaller in order to avoid chip edges and extra sources. 

For all \xmm{} observations, we filtered out times with high background, defined as when the count rate in the 10-12 keV range was $>\mathrm{0.4\:cts\:s^{−1}}$ for the PN and $\mathrm{>0.35\;cts\; s^{−1}}$ for the MOS cameras.  The first observation suffered from a large background flare, with count rates of up to 18 cts $\mathrm{s^{-1}}$ in the PN camera, and 7-8 cts $\mathrm{s^{-1}}$ in the MOS cameras. The flare continued for longer in the MOS cameras, such that while only $\mathrm{45\%}$ of the PN exposure time was lost due to background flaring, $\mathrm{78\%}$ of the MOS1 exposure time and $\mathrm{85\%}$ of the MOS2 exposure time was lost to background flaring. In addition to background flaring, a known flare star (2MASS J16330429+2349464) was present 30\arcsec{} away from the quasar in the PN image. To avoid the star, we extracted the PN, MOS1, and MOS2 spectra using 20\arcsec{} radius source regions that did not include the star.

The second observation had discrete background flares instead of the overall high levels of the first observation, resulting in less time lost to background flares. Specifically, 30\% of the PN exposure time, 2\% of the MOS1 exposure time, and 3\% of the MOS2 exposure time were lost to background flaring. The flare star appeared in all three cameras even after background flare filtering. To avoid activity from the flare star, we also extracted the PN, MOS1, and MOS2 spectra of the second \xmm{} observation using 20\arcsec{} radius source regions.

\subsection{\chandra{}}
For \nineteenfivetwoeight{}, there were no soft X-ray observations contemporaneous with the \nustar{} observation, so we used the earlier \chandra{} data reported in SA20 (ObsID:\dataset[19528]{https://doi.org/10.25574/19528}). We used the spectrum from that paper, grouped to have a minimum of 1 count per bin. This was done using CIAO version 4.10 with CALDB version 4.8.0. The spectrum was extracted with a circular source region 2\arcsec{} in radius, with an annular background region centered on the source of inner radius 10\arcsec{} and outer radius 20\arcsec{}. We used energies 0.5 - 8.0 keV for the spectral fitting.

\begin{deluxetable*}{lcccccc}[b!]
\tablecaption{Target sample and observation details. \label{tab:obstable}}

\tablecolumns{5}
\tablewidth{0pt}
\tablehead{
\colhead{Target} &
\colhead{Observatory} &
\colhead{ObsID} &
\colhead{Date} & 
\colhead{Net Exposure Time} &
\colhead{Net Count Rate}\\ 
\colhead{} &
\colhead{} &
\colhead{} &
\colhead{} &
\colhead{(ks)} & 
\colhead{(cts ${\rm ks}^{-1}$)}
}
\startdata
\nineteenfivetwoeight{} & \chandra{} & 19528 & 2017-04-28 & 7.6 & 267.5\\ 
 {} & \nustar{} & 60601029002 & 2021-05-16 & 21.5/21.3 & 44.2/40.0\\ 
\PG & \swift{} & 00089146001 & 2021-06-08 & 1.7 & 144.7\\ 
{} & \nustar{} & 60601030002 & 2021-06-08 & 36.5/36.3 & 68.1/65.7\\ 
\FBQS & \xmm{} &  0870910101 & 2020-08-08 & 57.6/5.8/3.6 & 537.7/118.5/146\\ 
{} & {} & 0870910301 & 2021-01-31 & 73.49/199.4/103.8 & 748.4/199.9/195.6\\ 
{} & \nustar{} & 60601012002 & 2020-08-09 & 102.2/101.2 & 21.2/19.5\\ 
{} & {} & 60601012004 & 2021-01-31 & 101.1/100.0 & 22.9/20.8 \\& 
\enddata
\tablecomments{\nustar{} net exposure times and net count rates are written as FPMA/FPMB. \xmm{} net exposure times and net count rates are written as PN/MOS1/MOS2. }
\end{deluxetable*}

\section{X-ray Properties}

\subsection{Average Spectra}
\begin{figure}
    \plotone{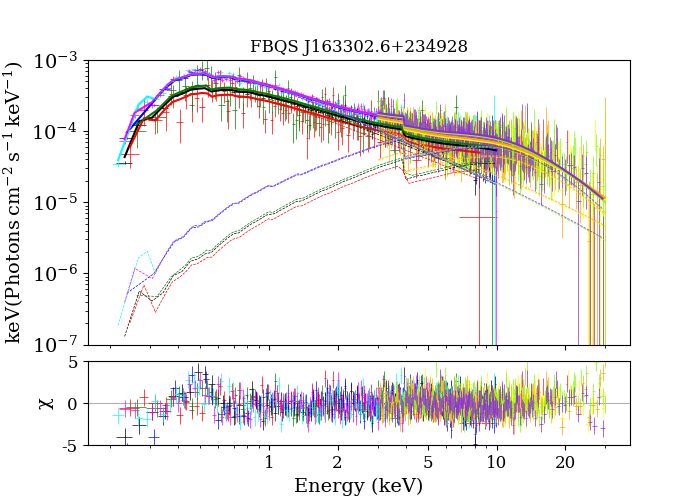}
    \caption{Unfolded spectrum and best-fit TBABS*ZPHABS*(CUTOFFPL+PEXRAV) model for \FBQS{}. Black, red and green correspond to PN, MOS1, and MOS2 data from \xmm{} observation 0870910101. Blue, cyan, and magenta correspond to PN, MOS1 and MOS2 data from \xmm{} observation 0870910301. Yellow and orange correspond to FPMA and FPMB data from \nustar{} observation 60601012002.  Chartreuse and purple correspond to FPMA and FPMB data from \nustar{} observation 60601012004. The model plotted is the fit reported in Table \ref{tab:FBQS_varying_pex_params}.}
    \label{fig:FBQS_uf}
\end{figure}

\begin{figure}
    \plotone{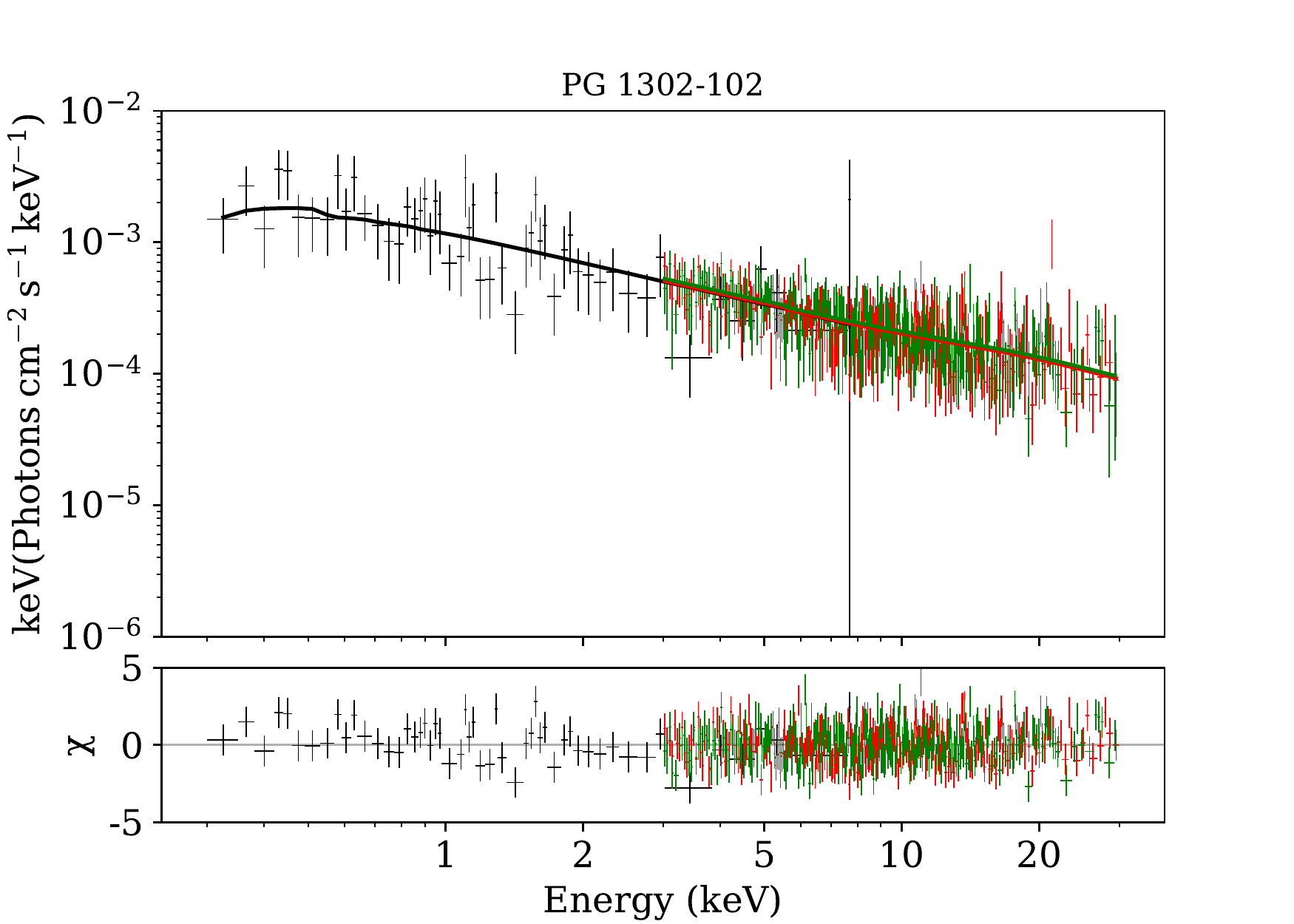}
    \caption{Unfolded spectrum and best-fit {\bf TBABS*ZPHABS*CUTOFFPL model} for \PG{}. Black corresponds to \swift{} XRT data; red and green correspond to \nustar{} FPMA and FPMB data. The model plotted is the fit reported in Table \ref{tab:bestfit_params}.}
    \label{fig:PG1302_uf}
\end{figure}

\begin{figure}
    \plotone{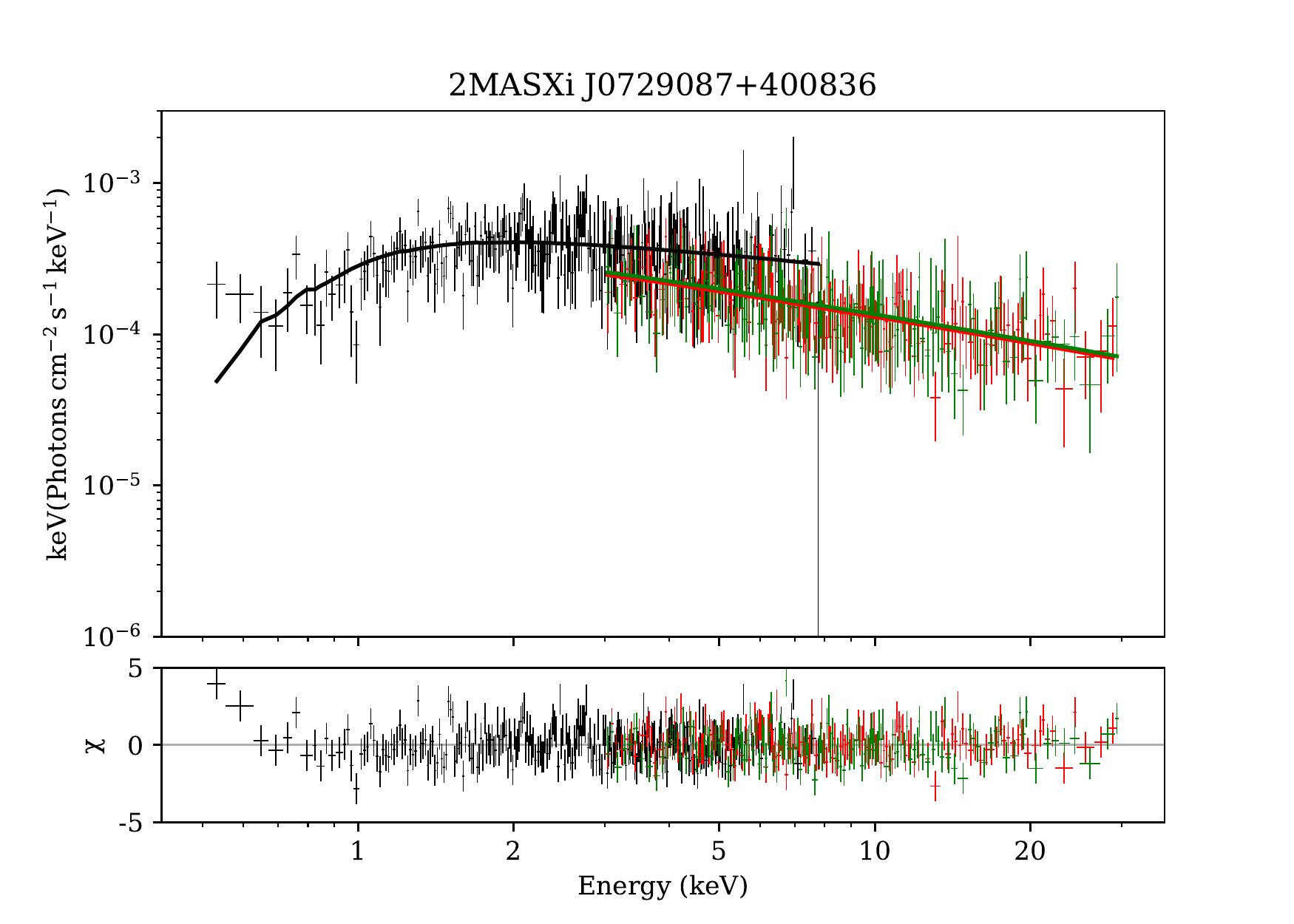}
    \caption{Unfolded spectrum and best-fit TBABS*ZPHABS*CUTOFFPL model for \nineteenfivetwoeight{}. Black corresponds to \chandra{} data; red and green correspond to \nustar{} FPMA and FPMB data. The model plotted is the fit reported in Table \ref{tab:2MASXi_varying_abs_params}.}
    \label{fig:2MASXi_uf}
\end{figure}

The unfolded spectra of the AGN are shown  as the error bars in Figures \ref{fig:FBQS_uf}, \ref{fig:PG1302_uf}, and \ref{fig:2MASXi_uf}. We  first considered average spectra. We fit all three  average spectra with a CONSTANT*TBABS*ZPHABS*CUTOFFPL model, with the CONSTANT term representing the cross-normalization constant between multiple observations, TBABS representing photoelectric absorption within our Galaxy, ZPHABS representing photoelectric absorption in the host galaxy, and the CUTOFFPL model representing the intrinsic spectrum of the black hole corona, which radiates approximately as a powerlaw spectrum  with an exponential cutoff. To take into account reflection components, we also tried a CONSTANT*TBABS*ZPHABS*(CUTOFFPL+PEXRAV) model, where the PEXRAV \citep{1995MNRAS.273..837M} parameter $R$ was set to be less than zero to ensure it represented the reflection component only. We fixed the inclination to be 30 degrees, the CUTOFFPL norm to be equal to the PEXRAV norm, and the CUTOFFPL Gamma to be the PEXRAV Gamma. This left $R$ as the only free parameter in the fit. \FBQS{} in particular appears to have a strong reflection component and its fit substantially benefits from the addition of a PEXRAV component. In contrast, \nineteenfivetwoeight{} and \PG{} did not show much improvement in C-stat/d.o.f.\ with the addition of a PEXRAV commponent. For this reason, we use the CONSTANT*TBABS*ZPHABS*CUTOFFPL fits as the best fits for these two sources, while we use the fit with PEXRAV for \FBQS{}. The best fits to the average spectra, along with their C-stat/d.o.f.\ and the observed \nustar{} FPMA fluxes, are presented in Table \ref{tab:bestfit_params}. The average spectra are well-fit by the {best-fit models} model, with C-stat/d.o.f.\ values $\approx$1, though \nineteenfivetwoeight{} shows some soft excess above the powerlaw component, a feature found in most AGN spectra below about 2~keV \citep{2020MNRAS.496.4255B}. The physical origin of this soft excess is not clear but has been proposed to be 
smeared reflected emission lines \citep{2006MNRAS.365.1067C,2013MNRAS.428.2901W} or a warm ($\mathrm{\sim 0.1\, keV}$) component of the corona \citep{2011A&A...534A..39M,2012MNRAS.420.1848D}. \nineteenfivetwoeight{} shows $\sim$ 40\% variability between its \chandra{} and \nustar{} observations. \FBQS{} shows $\sim$ 50\% variability between its two \xmm{} observations, and 7\% variability between its two \nustar{} observations. 

From the best fit XSPEC models, we measure the average spectral index, $\mathrm{\Gamma}$, for the three AGN. Taking into account the \nustar{} data, the values are softer than measured in the \chandra{} data alone in SA20. This is particularly true for \FBQS{}, likely because of the addition of the reflection component compared to the fit from SA20. There is overlap between the 90\% confidence intervals for $\mathrm{\Gamma}$ from SA20 and this paper for \nineteenfivetwoeight{} and \PG{}, meaning these measurements are consistent with each other. The same is not true for \FBQS{}'s value of $\mathrm{\Gamma}$, which does not overlap with its 90\% confidence interval from SA20. \PG{}'s $\mathrm{\Gamma}$ is within the 1.5-2.0 range typical for AGN \citep{1994MNRAS.268..405N,2008ApJ...682...81S,2013MNRAS.433.2485B}.  However \nineteenfivetwoeight{} has slightly harder values of $\mathrm{\Gamma}$ than the typical range, while \FBQS{} has a slightly softer value of $\mathrm{\Gamma}$ than typical. The value of the PEXRAV $R$ for \FBQS{} is extremely high ($7.67^{+1.38}_{-1.30}$), much larger than the typical 1-2 range for AGN, implying a very hard spectrum.  Similarly high values, however, were found for a few sources in \citet{2017ApJS..233...17R}; e.g., Mrk~1310 is best fit with $R=6.7$ and ESO~438-9 is best fit with $R=7.8$. We attempted to set an upper limit of 2 on $R$ and refit, to see if the value of $\mathrm{\Gamma}$ would also become hard. However, the value of $\mathrm{\Gamma}$ only went down to $2.12^{+0.02}_{-0.01}$, which is still soft. We also tried ignoring the energies below 2~keV to avoid the possibility of any soft excess biasing the value of $\mathrm{\Gamma}$. In this case, $\mathrm{\Gamma}$ went down only to $2.02\pm0.10$, which is still soft.

We compare the spectral indices of our sample from the average spectral fits to the spectral indices of the BAT AGN Spectroscopic Survey (BASS) sample \citep{2017ApJS..233...17R}. This sample includes 838 AGN. We exclude the blazars from the sample as they have different X-ray spectra than non-blazar AGN. Excising the blazars, the sample contains 703 AGN at redshift $0.001\leq z  \leq 0.65$, with bolometric luminosities ${39.54 \leq {\rm log}(L_{\rm bol} / {\rm erg}\, {\rm s}^{-1}) \leq 47.75}$.  The BASS sample properties envelope the three targets discussed here. The $\mathrm{\Gamma}$ values for our sample measured using the TBABS*ZPHABS*(CUTOFFPL+PEXRAV) average spectral fits  are compared to the BASS $\mathrm{\Gamma}$ values in their Table 5. 
The BASS values of $\mathrm{\Gamma}$ were measured using a PEXRAV component with a CUTOFFPL input freely fit; therefore they take into account both an intrinsic continuum and reflection component. We use the PEXRAV-containing fits of our three AGN for comparison even though they are not required for \nineteenfivetwoeight{} and \PG{}. This is in order to be consistent with the BASS spectral model and to use our full sample in the comparison.  We performed a Kolmogorov-Smirnoff test comparing the three candidate binary SMBH sources grouped together as a distribution against the BASS sample, making a cut to the BASS sample based on Eddington ratio. For the BASS sample, we use the Eddington ratios from \citet{2017ApJ...850...74K}, preferentially using values calculating based on \swift{} 14-195 keV luminosity, when available. In cases where this was not available, we used the Eddington ratios calculated from the 5100 \AA{} luminosity. A total of 319 AGN in the BASS sample have Eddington ratios measured in these two ways. We restricted the Eddington ratio to be ${-1.08\leq {\rm log}(L/L_{\rm Edd})\leq -0.12}$, leaving a sample of 170 AGN to compare to our three AGN with a KS test.  The resulting p-value of 0.867 is too high to reject the null hypothesis that our three candidate binary SMBH AGN are drawn from the same distribution as the BASS sample.  This is the same conclusion we came to in SA20 for a larger sample with only soft X-ray data.

We then investigated how the sample compares to the general AGN population when including \nustar{} data above 10~keV.  For this comparison we compared the average spectral indices of our sample against the sample of 195 unobscured AGN in \citet{2022ApJ...927...42K}. These AGN are a subset of the BASS sample and have \nustar{} observations in addition to \swift{} XRT and \xmm{} observations. We cross matched Table 1 in \citet{2022ApJ...927...42K} with the BAT 70-month survey \citep{2013ApJS..207...19B} as well as \citet{2017ApJS..233...17R} to retrieve 2-10 keV luminosities, and \citet{2017ApJ...850...74K} to retrieve Eddington ratios. \citet{2022ApJ...927...42K} fit the AGN with models including just a CUTOFFPL component, models with a CUTOFFPL+PEXRAV component, and models with an XILLVERCP component. We performed a KS test to compare our measured spectral indices to the spectral indices in Table 1 of \citet{2022ApJ...927...42K}, specifically the ones measured using a phenomenological reflection (PEXRAV) model. There were a total of 103 AGN with $\mathrm{\Gamma}$ values measured using the PEXRAV-containing model. The 103 AGN with the PEXRAV model range from 
$0.002\leq z  \leq 0.197$, with bolometric luminosities ${42.44 \leq {\rm log}(L_{\rm bol} / {\rm erg}\, {\rm s}^{-1}) \leq 46.61}$. Multiple fits were listed for some AGN, so where there were duplicates, we chose the fit with the C-stat/d.o.f value closest to one. We find that 67 of the AGN with $\mathrm{\Gamma}$ values had Eddington ratios listed in \citet{2017ApJ...850...74K}. We made the same cut on Eddingtion ratio as we did for our test versus the BASS sample, ${-1.08\leq {\rm log}(L/L_{\rm Edd})\leq -0.08}$. This left a total of 49 AGN in the sample. After performing the KS test grouping our sample as a distribution versus the \citet{2022ApJ...927...42K} sample, the resulting p-value was 0.990.We cannot reject the null hypothesis that  our values of $\mathrm{\Gamma}$ are drawn from the same distribution as the $\mathrm{\Gamma_{PEX}}$ values of \citet{2022ApJ...927...42K}.

It is worth comparing the very high value of $R$ of \FBQS{} to the values of $R$ of the BASS sample as well. When compared to all non-blazar AGN in the BASS sample with measured $R$ values (184 AGN), there are only two AGN with $R$ values greater than or equal to the 90\% confidence interval for \FBQS{}, namely Mrk~1310 with $R=6.7$ and ESO~438-9 with $R=7.8$.

\begin{figure}
\plotone{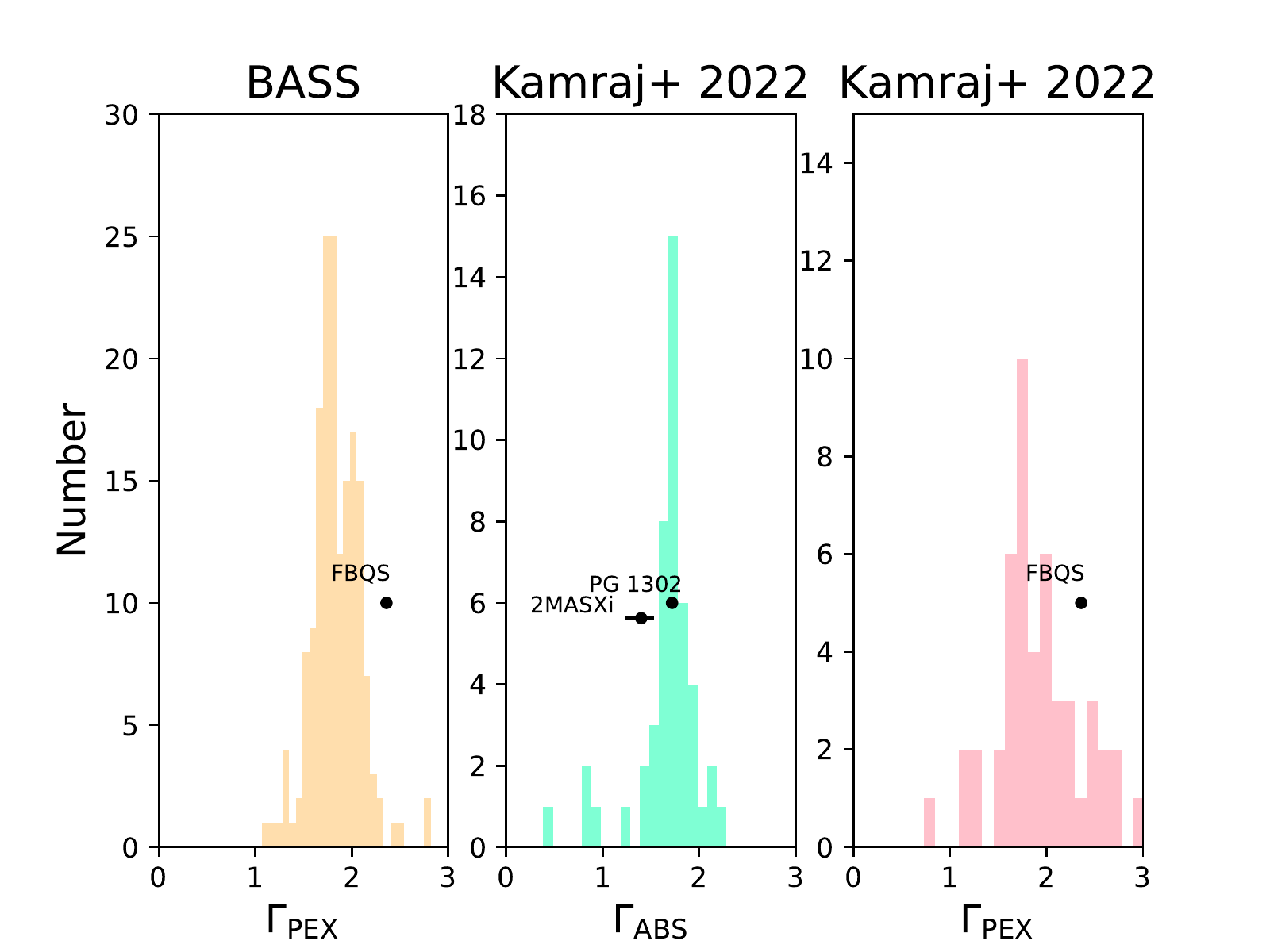}
    \caption{Spectral indices  derived from average spectra for the three AGN in our sample compared to histograms of the comparison samples used for the KS test. The histograms themselves depict $\mathrm{\Gamma}$ values from the BASS sample \citep{2017ApJS..233...17R} (left panel),   $\mathrm{\Gamma_{ABS}}$ values from \citet{2022ApJ...927...42K} (middle panel), and $\mathrm{\Gamma_{PEX}}$ values from \citet{2022ApJ...927...42K} (right panel).  The leftmost panel and rightmost panel use the $\Gamma$ from the average TBABS*ZPHABS*(CUTOFFPL+PEXRAV) fit of \FBQS{}. The middle panel uses $\mathrm{\Gamma}$ from the average TBABS*ZPHABS*CUTOFFPL fit of \nineteenfivetwoeight{} and \PG{}. In all three cases, the candidate binary SMBH AGN are not substantially distinct from the larger general AGN populations.}
    \label{fig:hist}
\end{figure}

At energies below $\sim 1$ keV, emission from the viscously heated circumbinary disc dominates, while the higher energy spectrum is dominated by shock-heated gas in the minidiscs around each SMBH, streams, and near the cavity wall at the inner edge of the circumbinary disk.  The presence of the cavity could lead to a small depression at a few keV \citep[e.g.,][]{2018MNRAS.476.2249T} where the cavity in the circumbinary disk cuts out a range of temperatures. However, as explained in \citet{2018MNRAS.476.2249T}, their simulations are initialized with an artificially high disk temperature for numerical reasons.  A single-BH disk with similar parameters, experiencing only viscous heating, would be 10-100 times cooler, moving the notch to correspondingly lower photon energies.  Additional heating by the binary, through shocks, could maintain a high disk temperature, and cause a depression in the X-ray region, but future work is required to assess this. In any case, as seen in the unfolded X-ray spectra presented in Figures 1-3, there is no evidence of notches in the high-energy spectra of any of the sources.  This could be due to the notch signature being too subtle for the signal-to-noise ratio of our data, or due to the sources not in fact being binary SMBH AGN.  Alternatively,  theoretical work based on viscous heating of a standard geometrically thin, optically thick single-BH disk (i.e. without shock-heating due to a binary), predicts the notch to be at much lower energies, in the UV/optical/IR range \citep{2014ApJ...785..115R,2019ApJ...879..110K}. The shape and depth of the notch is also highly dependent on the binary parameters of the system. For example, \citet{2014ApJ...785..115R} express the temperature of the thermal emission that would be missing because of the notch in their equation 2, \begin{equation}{T_{0}=3.3\times 10^{4}[\dot{m}(\eta/0.1)^{-1} M_{8}^{-1} (a/100 R_{g})^{-3}]^{1/4}\; {\rm K}}
\end{equation} where ${\dot{m}}$ is the accretion rate in Eddington units, $\mathrm{\eta
 }$ is the radiative efficiency and $M_{8}$ is the black hole mass in units of $10^{8} M_{\odot}$. Using the mass and binary separations of of our sample (listed in Table 3 of SA20), the Eddington ratios listed in Table \ref{tab:prop_table} of this  paper, assuming that the binary separation is $a$, and assuming a radiative efficiency of 0.1, the resulting temperatures are 2,334 K for \nineteenfivetwoeight{}, 6,836 K for \PG{}, and 10,547 K for \FBQS{}. This would place the notch in the optical part of the spectrum for \nineteenfivetwoeight{}, and the ultraviolet part of the spectrum for \PG{} and \FBQS{}.

We also tested fits to these AGN with two powerlaw components representing the continuum instead of a single powerlaw component, under the assumption that if two black holes were present, their coronae would not necessarily have the same spectral indices. As expected for the additional degree of freedom, we did get improvements in the C-stat/d.o.f, but the powerlaw invariably became extremely soft, much steeper than the expectations of coronal emission.  This suggested the additional parameter was picking up a soft excess component of the spectrum, rather than a second corona. We added a third powerlaw to the fits, but they did not improve C-stat/d.o.f.  We therefore do not find any evidence of two coronae in these data.

\subsection{Spectral and Flux Variability}


There is the potential for spectral variability between the different epochs of observation of \nineteenfivetwoeight{} and \FBQS{}. For \FBQS{} we tested fits where ZPHABS $N_\mathrm{H}$, CUTOFFPL $\mathrm{\Gamma}$ and $E_\mathrm{cut}$, and PEXRAV $R$ varied. We found that including variability in $\mathrm{\Gamma}$ and $R$ improved the fits, while including variability in the others did not. We tabulate the results of the best TBABS*ZPHABS*(CUTOFFPL+PEXRAV) fit with $\mathrm{\Gamma}$ and $R$ free to vary in Table \ref{tab:FBQS_varying_pex_params}. The latter is also plotted as the model in Figure \ref{fig:FBQS_uf}. The value of $\mathrm{\Gamma}$ in these fits becomes softer when the source is brighter, which is consistent with the well-established correlation between $\mathrm{\Gamma}$ and the Eddington ratio \citep[e.g.,][]{2013MNRAS.433.2485B}. During the first epoch, $R$ is more in line with the range of $R$ values seen in the BASS sample, with 14 AGN possessing $R$ values within its 90\% confidence interval, plus two that exceed it. However the second epoch $R$ reached an even higher value than in the average fit, $8.83^{+1.56}_{-1.37}$. Only one BASS AGN has an $R$ value within this 90\% confidence interval. As with the average spectrum, we set an upper limit of 2 on $R$ and refit to see if the value of $\mathrm{\Gamma}$ would become harder, but it only went down to $2.04\pm0.03$ in the first epoch and $2.14\pm0.02$ in the second epoch. We also tried ignoring energies <2 keV, but in this case $\mathrm{\Gamma}$ was only $2.03^{+0.15}_{-0.14}$ in the first epoch and $2.01\pm0.11$ in the second epoch, which are still soft values.

For \nineteenfivetwoeight{}, we found that allowing $\mathrm{\Gamma}$ to vary between epochs improved the fit, while allowing the other spectral parameters to vary did not. The best TBABS*ZPHABS*CUTOFFPL fit with varying $\mathrm{\Gamma}$ is tabulated in Table  \ref{tab:2MASXi_varying_abs_params}. The source is harder in the \chandra{} observation than in its \nustar{} observations, also consistent with the prevailing trend of harder $\mathrm{Gamma}$ values with increasing Eddington ratio. Since the 90\% confidence intervals of $\mathrm{\Gamma}$ overlap between the two epochs, the evidence for spectral variability in this source is weaker than for \FBQS{}.

\PG{} had a single set of simultaneous \swift{} and \nustar{} observations, so we did not attempt to fit it with a model that had parameters vary between the observations. We plot its average spectrum in Figure \ref{fig:PG1302_uf} using the fit from Table \ref{tab:bestfit_params}.

The rest-frame 2-10 keV fluxes for the AGN in our sample were measured from the CUTOFFPL component in \nineteenfivetwoeight{} and \PG{}, and from the sum of the CUTOFFPL and PEXRAV components in \FBQS{}. From these we derived the rest frame 2-10 keV luminosities using the luminosity distances listed in NED. These are tabulated in Table \ref{tab:prop_table}. We then calculated the bolometric luminosities using the rest frame 2-10 keV luminosity and the universal expression for the bolometric correction ${K_{X}(L_{X})}$ from Table 1 of \citet{fred}. Finally, the bolometric luminosities were divided by the Eddington luminosity for each object, which was estimated using $L_{\rm Edd}=1.26 \times 10^{38}\, (M_{\rm BH} / M_{\odot})\, {\rm erg}\, {\rm s}^{-1}$. The SMBH masses were derived from Table 3 of SA20. The rest frame 2-10 keV luminosities and the Eddington ratios are also listed in Table \ref{tab:prop_table}. For \nineteenfivetwoeight{} and \FBQS{}, we include the epoch-by-epoch fluxes, luminosities, and Eddington ratios as separate lines in Table~\ref{tab:prop_table}.

\begin{deluxetable*}    
{lccccccc}
\tablecaption{Parameters for best-fit average models.\label{tab:bestfit_params}}
\tablewidth{2pt}
\tablehead{ \colhead{} & \colhead{ZPHABS} &  \multicolumn{3}{c}{CUTOFFPL} &  \colhead{PEXRAV} & \colhead{} & \colhead{}\\ 
\cline{3-5}   \colhead{Target} & \colhead{$N_{\rm H}$} &  \colhead{$\mathrm{\Gamma}$} & \colhead{$E_{\rm cut}$} & \colhead{Norm} & \colhead{$R$} & \colhead{C-stat/d.o.f.} & \colhead{Observed Flux}\\
\colhead{} &  \colhead{$\mathrm{(10^{22}\;cm^{-2})}$} & \colhead{} & \colhead{(keV)} & \colhead{($10^{-4}$ cts $\mathrm{s^{-1}\:keV^{-1}}$)} & \colhead{} & \colhead{} &  \colhead{$\mathrm{(10^{-13}\;erg\;cm^{-2}\;s^{-1})}$}}
\startdata
\nineteenfivetwoeight{} & $0.24^{+0.10}_{-0.09}$ & $1.40^{+0.13}_{-0.16}$ & {$>30.05$} & $3.85^{+0.84}_{-0.69}$ & N/A & {1036.67/1143}  & {$21.9\pm{1.0}$}\\
\PG{} & {<0.01} & {$1.72\pm{0.05}$} & {$>186.62$} & {$10.98^{+1.11}_{-1.01}$} & N/A & {1103.37/1121}  & {$39.8^{+1.9}_{-1.8}$}\\
\FBQS{} & {$0.14\pm{0.01}$} & {$2.36\pm{0.05}$} & {$66.73^{+15.73}_{-10.94}$} & {$4.64^{+0.17}_{-0.16}$} & {$7.67^{+1.38}_{-1.30}$} & {3114.70/2914} &  {$8.51^{+0.20}_{-0.19}$, $9.36^{+0.57}_{-0.55}$}
\enddata
\tablecomments{Error bars represent 90\% confidence intervals. For \nineteenfivetwoeight{} and \PG{} a TBABS*ZPHABS*CUTOFFPL model provided a satisfactory fit; for FBQS{}, a TBABS*ZPHABS*(CUTOFFPL+PEXRAV) fit was necessary. In fits with PEXRAV components, abundances were  set to solar, the PEXRAV cutoff energy was set equal to the CUTOFFPL cutoff energy, and the PEXRAV norm was set equal to the CUTOFFPL norm, leaving $R$ as the only free PEXRAV parameter. For observed fluxes, the \nustar{} FPMA flux of the best average fit was used. The \chandra{} normalization constant value for \nineteenfivetwoeight{} was $1.71^{+0.16}_{-0.15}$ (ObsID: 19528). The \swift{} XRT normalization constant for \PG{} was $1.08^{+0.15}_{-0.14}$ (ObsID: 00089146001). The \xmm{} normalization constants for the first observation of \FBQS (ObsID: 0870910101) were $0.77\pm{0.03}$ for PN, $0.67\pm{0.05}$ for MOS1, and $0.83^{+0.07}_{-0.06}$ for MOS2. The \xmm{} normalization constants for the second observation of \FBQS{} (ObsID: 0870910301) were $1.11\pm{0.04}$ for PN, $1.12\pm{0.04}$ for MOS1, and $1.13\pm{0.04}$ for MOS 2. The \nustar{} FPMA constant for the second observation of \FBQS{} (ObsID: 60601012004) was $1.14\pm{0.04}$.}
\end{deluxetable*}

\begin{deluxetable}{@{\extracolsep{10pt}}lcccccl@{}}
\tablecaption{Parameters for best TBABS*ZPHABS
(CUTOFFPL+PEXRAV) fit with varying spectral parameters for \FBQS{}}\label{tab:FBQS_varying_pex_params}
\tablewidth{2pt}
\tablehead{ \colhead{} & \colhead{ZPHABS} &  \multicolumn{3}{c}{CUTOFFPL} &  \colhead{PEXRAV} & {}\\ 
\cline{3-5}   \colhead{Obs. Date} & \colhead{$N_{\rm H}$} &  \colhead{$\mathrm{\Gamma}$} & \colhead{$E_{\rm cut}$} & \colhead{Norm} & \colhead{$R$} & \colhead{C-stat/d.o.f.}\\
\colhead{} &  \colhead{$\mathrm{(10^{22}\;cm^{-2})}$} & \colhead{} & \colhead{(keV)} & \colhead{($10^{-4}$ cts $\mathrm{s^{-1}\:keV^{-1}}$)} & \colhead{} & \colhead{}}
\startdata
{2020-08-08} & $0.14\pm{0.01}$ & $2.22\pm{0.05}$ & $64.26^{+15.10}_{-10.46}$ & $4.54^{+0.23}_{-0.22}$ & $4.97^{+1.22}_{-1.02}$ & 3015.037/2912\\
2021-01-31 & {} & $2.40\pm{0.05}$ & {} & {} & $8.83^{+1.56}_{-1.37}$ &  {}\\
\enddata
\tablecomments{Error bars represent 90\% confidence intervals. Abundances were  set to solar, the PEXRAV cutoff energy was set equal to the CUTOFFPL cutoff energy, and the PEXRAV norm was set equal to the CUTOFFPL norm, leaving $R$ as the only free PEXRAV parameter.  The \xmm{} normalization constants for the first observation of \FBQS (ObsID: 0870910101) were $0.80^{+0.04}_{-0.03}$ for PN, $0.69\pm{0.05}$ for MOS1, and $0.86^{+0.08}_{-0.07}$ for MOS2. The \xmm{} normalization constants for the second observation of \FBQS{} (ObsID: 0870910301) were $1.12\pm{0.05}$ for PN, $1.15\pm{0.06}$ for MOS1, and $1.15\pm{0.06}$ for MOS 2. The \nustar{} FPMA constant for the second observation of \FBQS{} (ObsID: 60601012004) was $1.18\pm{0.07}$.
}
\end{deluxetable}

\begin{deluxetable}{@{\extracolsep{10pt}}lccccc@{}}
\tablecaption{Parameters for best TBABS*ZPHABS*CUTOFFPL fit with varying spectral parameters for \nineteenfivetwoeight{}.}\label{tab:2MASXi_varying_abs_params}
\tablewidth{2pt}
\tablehead{ \colhead{} & \colhead{ZPHABS} & \multicolumn{3}{c}{CUTOFFPL} & \colhead{}\\ 
\cline{3-5}   \colhead{Obs. Date} & \colhead{$N_{\rm H}$} & \colhead{$\mathrm{\Gamma}$} & \colhead{$E_{\rm cut}$} & \colhead{Norm} & \colhead{C-stat/d.o.f.}\\
\colhead{} &  \colhead{$\mathrm{(10^{22}\;cm^{-2})}$} & \colhead{} & \colhead{(keV)} & \colhead{($10^{-4}$ cts $\mathrm{s^{-1}\:keV^{-1}}$)} & \colhead{}
}
\startdata
2017-04-28 & {$0.27^{+0.20}_{-0.09}$} & {$1.33^{+0.23}_{-0.06}$} & {$>69.5$} & {$4.81^{+0.89}_{-0.83}$} & {1030.36/1143}\\
2021-05-16 & {} & $1.51^{+0.13}_{-0.08}$ & {} & {} & {}\\
\enddata
\tablecomments{Error bars represent 90\% confidence intervals. The \chandra{} normalization constant for the first observation of \nineteenfivetwoeight{}{} (ObsID: 19528) is $1.19^{+0.30}_{-0.23}$.}
\end{deluxetable}


\section{Discussion}
\citet{2014ApJ...785..115R} predict that binary SMBH systems will have excess shock-heated gas with temperature on the order $\sim$ 100 keV for a separation of $\sim 100\, r_{g}$, where the gravitational radius $r_{g} \equiv GM/c^2$, $G$ is the gravitational constant, $M$ is the total binary mass, and $c$ is the speed of light. This gas is heated by the impact of narrow streams of accreting material as they fall from the circumbinary disk into the mini-disks around the individual black holes. This would lead to a thermal component to the X-ray spectrum separate from the standard coronal powerlaw.  Following \citet{2014ApJ...785..115R}, we can estimate the temperature of the shocked gas of the binary SMBH using their equation 12, i.e.
\begin{equation} {T_{s1,2} = 6.2 \times 10^{10} [(1.4 + 1.2Zm_e/m_p)/(1.1 + 1.2Z)] (a/100R_g)^{-1} \Phi^2_{1,2}(1+q)^{-1} (q^{0.3},q^{0.7})\, {\rm K}}.
\end{equation}
where the 1,2 notation denotes the primary or secondary black hole, and $Z$ is the ratio of leptons to protons. $\Phi$ is a factor of order unity, with $\Phi_{1}\simeq0.9$ for the primary, and $\Phi_{2}\simeq0.6$ for the secondary, assuming a binary mass ratio of 0.3. Using values for $a$ from SA20, the redshifts of the AGN, and assuming $q=0.3$ and $Z=1$, we can find observed frame energies for the shock-heated gas around each individual black hole. These energies are 78.6 keV (primary) and 21.6 keV (secondary) for \nineteenfivetwoeight{}, 275 keV (primary) and 75 keV (secondary) for \PG{}, and 2.48 MeV (primary) and 680.81 keV (secondary) for \FBQS{}. While most of these energies are outside the observational range of \nustar{}, the energy for the secondary of \nineteenfivetwoeight{} is within the detectable range. \citet{2014ApJ...785..115R} note, however, that these values are upper limits  because cooling via efficient electron/positron pair production could decrease the temperature of the shocked gas. This raises the possibility that the excess energy in \PG{} and \FBQS{}, as well as from the primary of \nineteenfivetwoeight{}, would be detectable in the X-ray data considered here. The uncertainty as to the excess's true location means we cannot definitely state whether we have detected it or not. Detecting the excess X-ray emissions due to shocked accretion streams would be difficult even if the temperatures were low enough to be in the soft X-ray or \nustar{} bands. A soft X-ray excess would resemble the soft excess seen in many single-SMBH AGN, while a hard X-ray excess would resemble the Compton reflection component also seen in many single-SMBH AGN. If the soft excess seen in single-SMBH AGN is blurred reflection, an observatory with high spectral resolution like \xrism{} \citep{2020arXiv200304962X} would likely be able to tell it apart from a thermal soft excess expected as a result of emission from the accretion streams and mini-disks.

The case of a hard excess from a binary mimicking an enhanced reflection component is particularly relevant for \FBQS{}, which shows an extremely high value of $R$ in both its average spectrum and especially the second epoch. This could potentially be evidence of excess hard X-ray emission from a binary SMBH. However it should be noted that even its high $R$ value is not entirely out of the range known from single SMBH AGN. In particular, Mrk 1310 and ESO 438-9 in the BASS sample show $R$ values of 6.7 and 7.8, respectively, and neither is known to be a binary SMBH candidate. Even during the second epoch, \FBQS{} is less than 3$\sigma$ away from ESO 438-9. So while the very hard spectrum of \FBQS{} is interesting, it cannot be conclusively stated to be outside the range of normal AGN.

The predicted binary separations for these sources (listed in Table 3 of SA20) are all too small to place the enhanced X-ray emission in the \nustar{} band, with the exception of the secondary's minidisk for \nineteenfivetwoeight{}.  While \nineteenfivetwoeight{} does have a slightly harder spectrum than typical for an AGN, the difference is not statistically significant. The absence of hardening in in \nineteenfivetwoeight{}'s spectra investigated in this paper potentially imply that cooling due to pair production is inefficient.

We have shown that broadband X-ray spectra, spanning the soft X-rays observed by \chandra, \swift, and \xmm{} to the hard X-rays observed by \nustar, do not exhibit any evidence that the three candidate binary SBMH AGN studied here are dramatically different from the typical (single) AGN population.  The broadband X-ray spectral indices are not distinct from larger AGN populations, despite some theoretical predictions that enhanced X-ray emission should be expected. We find no evidence for a notch in their X-ray spectra, nor do we find evidence for multiple X-ray coronae.  There are multiple potential explanations for this non-result. First, the signatures of a binary SMBH might be too subtle given the quality of our data.  In that case, deeper observations or future, more sensitive facilities might detect indications of binarity missed in the current data.  Second, as emphasized in SA20, theoretical predictions of high-energy emission from binary SMBH AGN are relatively immature as a field, with models still highly idealized.  In that case, observations such as these test current models and will help direct future theoretical modeling.

Finally, it is possible the three AGN investigated in this paper are not in fact binary SMBHs. Indeed, we know that not all binary SMBH candidates can truly be binaries, since this would overpredict the gravitational wave background seen by pulsar timing arrays \citep{2018ApJ...856...42S}. The literature is rife with claimed periodicity in AGN light curves, though follow-up analyses find many claims to be statistically lacking \citep[e.g.,][]{2016MNRAS.461.3145V, 2018ApJ...859...10B}. Recent work shows that \PG{}'s variations cannot be explained by random noise, and there is very strong support for periodicity or, at least, quasiperiodicity \citep{2020ApJ...900..117Z}, where the latter is an expected consequence of SMBH binaries \citep{2019ApJ...879...76B, 2022ApJ...928..187C}.  We also note that even if the periodicity (or quasiperiodicity) is real, it might not be due to a binary SMBH system. Similar to the mechanism that causes quasi-periodic oscillations (QPOs) in systems with a stellar mass compact object \citep[e.g.,][]{2020ApJ...900..117Z}, periodicity could potentially be due to precession of the accretion disk or jet \citep[e.g.,][]{2022arXiv220506275D}. However, as noted by \citet{2015MNRAS.453.1562G},  the timescale for a warped accretion disk (such that it would undergo precession) to remain before self-gravity undoes the warp is much shorter than the typical AGN lifetime. It is therefore unlikely that precession due to a warped accretion disk explains the behavior of the three AGN studied here.

Ongoing and future synoptic surveys will improve and extend the light curves for candidate periodic AGN, testing their unusual variability with increasing statistical accuracy. In particular, \citet{2019NewAR..8601525D} mention the Rubin Observatory will scan the entire observable optical sky every 3 days, and will be able to monitor 
$10^{4}-10^{5}$ AGN in the Deep Drilling Fields with an even higher cadence. It will be by far the best survey in terms of signal to noise, sampling, and duration for identifying and confirming periodic AGN candidates.

Some current candidates will likely fall as a result of additional monitoring, while new candidates will be identified.  X-ray emission, coming from the innermost regions around the SMBH(s), should, in principle, provide a strong test if observed periodicity is due to binarity. Our results, emphasizing soft X-ray data in SA20 and broadband X-ray data here, do not find evidence for unusual X-ray properties for some of the strongest and X-ray brightest binary SMBH candidates currently known.  However, we are still in the early stages of this field, both observationally and theoretically.  The non-detections reported here can help motivate future, more sensitive observations (and observatories), while simultaneously helping direct theoretical work.

\begin{deluxetable*}{lcccll}
\tablecaption{Table of AGN Properties.\label{tab:prop_table}}
\tablewidth{0pt}
\tablehead{
\colhead{Target} & 
\colhead{Obs Date} &
\colhead{$z$} &
\colhead{$f_{2-10}$\tablenotemark{a}} &
\colhead{$\mathrm{log}(L_{2-10})$\tablenotemark{b}} &
\colhead{$L / L_{\rm Edd}$\tablenotemark{c}}}
\startdata
\nineteenfivetwoeight{} & {avg.} & 0.074 & {$22.91^{+1.1}_{-1.0}$} & {$43.52\pm{0.02}$} & 0.08  \\
{} & {2017-04-28} & {} & {$23.4^{+0.6}_{-2.5}$} & {$43.50^{+0.04}_{-0.02}$} & {0.08}\\
{} & {2021-05-16} & {} & {$24.6\pm{1.7}$} &  {$43.55\pm{0.03}$} & {0.09}\\
\PG{} & {2021-06-08} & 0.278 & {$38.90^{+1.83}_{-0.89}$} & {$45.01\pm{0.01}$} & 0.76 \\
\FBQS{} & {avg.} & 0.821  & $8.51^{+0.20}_{-0.19}$ & {$45.46\pm{0.01}$} & 0.13 \\
{} & {2020-08-08} & {} & $9.12\pm0.21$ & $45.49\pm{0.01}$ & {0.15}\\
{} & {2021-01-31} & {} & $12.30^{+0.30}_{-0.20}$ & {$45.62\pm{0.10}$} & {0.20} 
\enddata
\tablenotetext{a}{Model flux from rest-frame 2-10 keV, in units of $\rm 10^{-13}\;erg\;cm^{-2}\;s^{-1}$.  Measured from CUTOFFPL component of TBABS*ZPHABS*CUTOFFPL fit for \nineteenfivetwoeight{} and \PG{}, from CUTOFFPL and PEXRAV components of  TBABS*ZPHABS*(CUTOFFPL+PEXRAV) fit for \FBQS{}.}
\tablenotetext{b}{Calculated using flux from previous column, in units of $erg\;s^{-1}$.}
\tablenotetext{c}{Calculated using bolometric correction $K_{X}(L_{X})$ from Table 1 of \citet{fred}. and the black hole masses listed in Table 3 of SA20.}\end{deluxetable*}

We thank Julian Krolik for helpful comments that have improved the paper. The scientific results reported in this article are based on data obtained from the \chandra{} Data Archive. This work is based on on observations obtained with \xmm{}, an ESA science mission with instruments and contributions directly funded by ESA Member States and NASA. We acknowledge the use of public data from the \swift{} data archive. The scientific results reported in this paper are based on data obtained from the \chandra{} Data Archive (\dataset[ObsID 19528]{https://doi.org/10.25574/19528}).  This research has
made use of data and/or software provided by the High Energy Astrophysics Science Archive Research Center (HEASARC), which
is a service of the Astrophysics Science Division at NASA/GSFC and the High Energy Astrophysics Division of the Smithsonian Astrophysical Observatory. This work has made use of data obtained from the \nustar{} mission, a project led by Caltech, funded by NASA and
managed by NASA/JPL. DJD received funding from the European Union's Horizon 2020 research and innovation program under Marie Sklodowska-Curie grant agreement No. 101029157, and from the Danish Independent Research Fund through Sapere Aude Starting Grant No. 121587. ZH acknowledges support from NASA ATP grant 80NSSC22K082.

\facilities{CXO, NuSTAR, XMM, Swift}

\software{HEASOFT \citep{2014ascl.soft08004N}, CIAO \citep{2006SPIE.6270E..1VF}, XMM-Newton SAS \citep{2004ASPC..314..759G}}

\bibliography{PeriodicQuasars2}{}

\begin{thebibliography}{}
\expandafter\ifx\csname natexlab\endcsname\relax\def\natexlab#1{#1}\fi
\providecommand{\url}[1]{\href{#1}{#1}}
\providecommand{\dodoi}[1]{doi:~\href{http://doi.org/#1}{\nolinkurl{#1}}}
\providecommand{\doeprint}[1]{\href{http://ascl.net/#1}{\nolinkurl{http://ascl.net/#1}}}
\providecommand{\doarXiv}[1]{\href{https://arxiv.org/abs/#1}{\nolinkurl{https://arxiv.org/abs/#1}}}

\bibitem[{{Amaro-Seoane} {et~al.}(2017){Amaro-Seoane}, {Audley}, {Babak}, {Baker}, {Barausse}, {Bender}, {Berti}, {Binetruy}, {Born}, {Bortoluzzi}, {Camp}, {Caprini}, {Cardoso}, {Colpi}, {Conklin}, {Cornish}, {Cutler}, {Danzmann}, {Dolesi}, {Ferraioli}, {Ferroni}, {Fitzsimons}, {Gair}, {Gesa Bote}, {Giardini}, {Gibert}, {Grimani}, {Halloin}, {Heinzel}, {Hertog}, {Hewitson}, {Holley-Bockelmann}, {Hollington}, {Hueller}, {Inchauspe}, {Jetzer}, {Karnesis}, {Killow}, {Klein}, {Klipstein}, {Korsakova}, {Larson}, {Livas}, {Lloro}, {Man}, {Mance}, {Martino}, {Mateos}, {McKenzie}, {McWilliams}, {Miller}, {Mueller}, {Nardini}, {Nelemans}, {Nofrarias}, {Petiteau}, {Pivato}, {Plagnol}, {Porter}, {Reiche}, {Robertson}, {Robertson}, {Rossi}, {Russano}, {Schutz}, {Sesana}, {Shoemaker}, {Slutsky}, {Sopuerta}, {Sumner}, {Tamanini}, {Thorpe}, {Troebs}, {Vallisneri}, {Vecchio}, {Vetrugno}, {Vitale}, {Volonteri}, {Wanner}, {Ward}, {Wass}, {Weber}, {Ziemer}, \& {Zweifel}}]{2017arXiv170200786A}
{Amaro-Seoane}, P., {Audley}, H., {Babak}, S., {et~al.} 2017, arXiv e-prints, arXiv:1702.00786.
\newblock \doarXiv{1702.00786}

\bibitem[{{Amaro-Seoane} {et~al.}(2023){Amaro-Seoane}, {Andrews}, {Arca Sedda}, {Askar}, {Baghi}, {Balasov}, {Bartos}, {Bavera}, {Bellovary}, {Berry}, {Berti}, {Bianchi}, {Blecha}, {Blondin}, {Bogdanovi{\'c}}, {Boissier}, {Bonetti}, {Bonoli}, {Bortolas}, {Breivik}, {Capelo}, {Caramete}, {Cattorini}, {Charisi}, {Chaty}, {Chen}, {Chru{\'s}li{\'n}ska}, {Chua}, {Church}, {Colpi}, {D'Orazio}, {Danielski}, {Davies}, {Dayal}, {De Rosa}, {Derdzinski}, {Destounis}, {Dotti}, {Du{\r{A}}{\textsterling}an}, {Dvorkin}, {Fabj}, {Foglizzo}, {Ford}, {Fouvry}, {Franchini}, {Fragos}, {Fryer}, {Gaspari}, {Gerosa}, {Graziani}, {Groot}, {Habouzit}, {Haggard}, {Haiman}, {Han}, {Istrate}, {Johansson}, {Khan}, {Kimpson}, {Kokkotas}, {Kong}, {Korol}, {Kremer}, {Kupfer}, {Lamberts}, {Larson}, {Lau}, {Liu}, {Lloyd-Ronning}, {Lodato}, {Lupi}, {Ma}, {Maccarone}, {Mandel}, {Mangiagli}, {Mapelli}, {Mathis}, {Mayer}, {McGee}, {McKernan}, {Miller}, {Mota}, {Mumpower}, {Nasim}, {Nelemans}, {Noble}, {Pacucci}, {Panessa}, {Paschalidis},
  {Pfister}, {Porquet}, {Quenby}, {Ricarte}, {R{\"o}pke}, {Regan}, {Rosswog}, {Ruiter}, {Ruiz}, {Runnoe}, {Schneider}, {Schnittman}, {Secunda}, {Sesana}, {Seto}, {Shao}, {Shapiro}, {Sopuerta}, {Stone}, {Suvorov}, {Tamanini}, {Tamfal}, {Tauris}, {Temmink}, {Tomsick}, {Toonen}, {Torres-Orjuela}, {Toscani}, {Tsokaros}, {Unal}, {V{\'a}zquez-Aceves}, {Valiante}, {van Putten}, {van Roestel}, {Vignali}, {Volonteri}, {Wu}, {Younsi}, {Yu}, {Zane}, {Zwick}, {Antonini}, {Baibhav}, {Barausse}, {Bonilla Rivera}, {Branchesi}, {Branduardi-Raymont}, {Burdge}, {Chakraborty}, {Cuadra}, {Dage}, {Davis}, {de Mink}, {Decarli}, {Doneva}, {Escoffier}, {Gandhi}, {Haardt}, {Lousto}, {Nissanke}, {Nordhaus}, {O'Shaughnessy}, {Portegies Zwart}, {Pound}, {Schussler}, {Sergijenko}, {Spallicci}, {Vernieri}, \& {Vigna-G{\'o}mez}}]{2023LRR....26....2A}
{Amaro-Seoane}, P., {Andrews}, J., {Arca Sedda}, M., {et~al.} 2023, Living Reviews in Relativity, 26, 2, \dodoi{10.1007/s41114-022-00041-y}

\bibitem[{{Arnaud}(1996)}]{1996ASPC..101...17A}
{Arnaud}, K.~A. 1996, in Astronomical Society of the Pacific Conference Series, Vol. 101, Astronomical Data Analysis Software and Systems V, ed. G.~H. {Jacoby} \& J.~{Barnes}, 17

\bibitem[{{Ballantyne} \& {Xiang}(2020)}]{2020MNRAS.496.4255B}
{Ballantyne}, D.~R., \& {Xiang}, X. 2020, \mnras, 496, 4255, \dodoi{10.1093/mnras/staa1866}

\bibitem[{{Barth} \& {Stern}(2018)}]{2018ApJ...859...10B}
{Barth}, A.~J., \& {Stern}, D. 2018, \apj, 859, 10, \dodoi{10.3847/1538-4357/aab3c5}

\bibitem[{{Baumgartner} {et~al.}(2013){Baumgartner}, {Tueller}, {Markwardt}, {Skinner}, {Barthelmy}, {Mushotzky}, {Evans}, \& {Gehrels}}]{2013ApJS..207...19B}
{Baumgartner}, W.~H., {Tueller}, J., {Markwardt}, C.~B., {et~al.} 2013, \apjs, 207, 19, \dodoi{10.1088/0067-0049/207/2/19}

\bibitem[{{Begelman} {et~al.}(1980){Begelman}, {Blandford}, \& {Rees}}]{1980Natur.287..307B}
{Begelman}, M.~C., {Blandford}, R.~D., \& {Rees}, M.~J. 1980, \nat, 287, 307, \dodoi{10.1038/287307a0}

\bibitem[{{Berczik} {et~al.}(2006){Berczik}, {Merritt}, {Spurzem}, \& {Bischof}}]{2006ApJ...642L..21B}
{Berczik}, P., {Merritt}, D., {Spurzem}, R., \& {Bischof}, H.-P. 2006, \apjl, 642, L21, \dodoi{10.1086/504426}

\bibitem[{{Bogdanovi{\'c}} {et~al.}(2022){Bogdanovi{\'c}}, {Miller}, \& {Blecha}}]{2022LRR....25....3B}
{Bogdanovi{\'c}}, T., {Miller}, M.~C., \& {Blecha}, L. 2022, Living Reviews in Relativity, 25, 3, \dodoi{10.1007/s41114-022-00037-8}

\bibitem[{{Bowen} {et~al.}(2019){Bowen}, {Mewes}, {Noble}, {Avara}, {Campanelli}, \& {Krolik}}]{2019ApJ...879...76B}
{Bowen}, D.~B., {Mewes}, V., {Noble}, S.~C., {et~al.} 2019, \apj, 879, 76, \dodoi{10.3847/1538-4357/ab2453}

\bibitem[{{Brightman} {et~al.}(2013){Brightman}, {Silverman}, {Mainieri}, {Ueda}, {Schramm}, {Matsuoka}, {Nagao}, {Steinhardt}, {Kartaltepe}, {Sanders}, {Treister}, {Shemmer}, {Brandt}, {Brusa}, {Comastri}, {Ho}, {Lanzuisi}, {Lusso}, {Nandra}, {Salvato}, {Zamorani}, {Akiyama}, {Alexander}, {Bongiorno}, {Capak}, {Civano}, {Del Moro}, {Doi}, {Elvis}, {Hasinger}, {Laird}, {Masters}, {Mignoli}, {Ohta}, {Schawinski}, \& {Taniguchi}}]{2013MNRAS.433.2485B}
{Brightman}, M., {Silverman}, J.~D., {Mainieri}, V., {et~al.} 2013, \mnras, 433, 2485, \dodoi{10.1093/mnras/stt920}

\bibitem[{{Caproni} {et~al.}(2013){Caproni}, {Abraham}, \& {Monteiro}}]{2013MNRAS.428..280C}
{Caproni}, A., {Abraham}, Z., \& {Monteiro}, H. 2013, \mnras, 428, 280, \dodoi{10.1093/mnras/sts014}

\bibitem[{{Charisi} {et~al.}(2016){Charisi}, {Bartos}, {Haiman}, {Price-Whelan}, {Graham}, {Bellm}, {Laher}, \& {M{\'a}rka}}]{2016MNRAS.463.2145C}
{Charisi}, M., {Bartos}, I., {Haiman}, Z., {et~al.} 2016, \mnras, 463, 2145, \dodoi{10.1093/mnras/stw1838}

\bibitem[{{Chen} {et~al.}(2020){Chen}, {Liu}, {Liao}, {Holgado}, {Guo}, {Gruendl}, {Morganson}, {Shen}, {Zhang}, {Abbott}, {Aguena}, {Allam}, {Avila}, {Bertin}, {Bhargava}, {Brooks}, {Burke}, {Carnero Rosell}, {Carollo}, {Carrasco Kind}, {Carretero}, {Costanzi}, {da Costa}, {Davis}, {De Vicente}, {Desai}, {Diehl}, {Doel}, {Everett}, {Flaugher}, {Friedel}, {Frieman}, {Garc{\'\i}a-Bellido}, {Gaztanaga}, {Glazebrook}, {Gruen}, {Gutierrez}, {Hinton}, {Hollowood}, {James}, {Kim}, {Kuehn}, {Kuropatkin}, {Lewis}, {Lidman}, {Lima}, {Maia}, {March}, {Marshall}, {Menanteau}, {Miquel}, {Palmese}, {Paz-Chinch{\'o}n}, {Plazas}, {Sanchez}, {Schubnell}, {Serrano}, {Sevilla-Noarbe}, {Smith}, {Suchyta}, {Swanson}, {Tarle}, {Tucker}, {Norbert Varga}, \& {Walker}}]{2020MNRAS.499.2245C}
{Chen}, Y.-C., {Liu}, X., {Liao}, W.-T., {et~al.} 2020, \mnras, 499, 2245, \dodoi{10.1093/mnras/staa2957}

\bibitem[{{Chen} {et~al.}(2022){Chen}, {Zhai}, {Liu}, {Guo}, {Peng}, {Li}, {SongSheng}, {Du}, {Hu}, \& {Wang}}]{2022arXiv220611497C}
{Chen}, Y.-J., {Zhai}, S., {Liu}, J.-R., {et~al.} 2022, arXiv e-prints, arXiv:2206.11497, \dodoi{10.48550/arXiv.2206.11497}

\bibitem[{{Combi} {et~al.}(2022){Combi}, {Lopez Armengol}, {Campanelli}, {Noble}, {Avara}, {Krolik}, \& {Bowen}}]{2022ApJ...928..187C}
{Combi}, L., {Lopez Armengol}, F.~G., {Campanelli}, M., {et~al.} 2022, \apj, 928, 187, \dodoi{10.3847/1538-4357/ac532a}

\bibitem[{{Crummy} {et~al.}(2006){Crummy}, {Fabian}, {Gallo}, \& {Ross}}]{2006MNRAS.365.1067C}
{Crummy}, J., {Fabian}, A.~C., {Gallo}, L., \& {Ross}, R.~R. 2006, \mnras, 365, 1067, \dodoi{10.1111/j.1365-2966.2005.09844.x}

\bibitem[{{d'Ascoli} {et~al.}(2018){d'Ascoli}, {Noble}, {Bowen}, {Campanelli}, {Krolik}, \& {Mewes}}]{2018ApJ...865..140D}
{d'Ascoli}, S., {Noble}, S.~C., {Bowen}, D.~B., {et~al.} 2018, \apj, 865, 140, \dodoi{10.3847/1538-4357/aad8b4}

\bibitem[{{De Rosa} {et~al.}(2019){De Rosa}, {Vignali}, {Bogdanovi{\'c}}, {Capelo}, {Charisi}, {Dotti}, {Husemann}, {Lusso}, {Mayer}, {Paragi}, {Runnoe}, {Sesana}, {Steinborn}, {Bianchi}, {Colpi}, {del Valle}, {Frey}, {Gab{\'a}nyi}, {Giustini}, {Guainazzi}, {Haiman}, {Herrera Ruiz}, {Herrero-Illana}, {Iwasawa}, {Komossa}, {Lena}, {Loiseau}, {Perez-Torres}, {Piconcelli}, \& {Volonteri}}]{2019NewAR..8601525D}
{De Rosa}, A., {Vignali}, C., {Bogdanovi{\'c}}, T., {et~al.} 2019, \nar, 86, 101525, \dodoi{10.1016/j.newar.2020.101525}

\bibitem[{{Done} {et~al.}(2012){Done}, {Davis}, {Jin}, {Blaes}, \& {Ward}}]{2012MNRAS.420.1848D}
{Done}, C., {Davis}, S.~W., {Jin}, C., {Blaes}, O., \& {Ward}, M. 2012, \mnras, 420, 1848, \dodoi{10.1111/j.1365-2966.2011.19779.x}

\bibitem[{{D'Orazio} {et~al.}(2015){D'Orazio}, {Haiman}, \& {Schiminovich}}]{2015Natur.525..351D}
{D'Orazio}, D.~J., {Haiman}, Z., \& {Schiminovich}, D. 2015, \nat, 525, 351, \dodoi{10.1038/nature15262}

\bibitem[{{Dorn-Wallenstein} {et~al.}(2017){Dorn-Wallenstein}, {Levesque}, \& {Ruan}}]{2017ApJ...850...86D}
{Dorn-Wallenstein}, T., {Levesque}, E.~M., \& {Ruan}, J.~J. 2017, \apj, 850, 86, \dodoi{10.3847/1538-4357/aa9329}

\bibitem[{{Dotti} {et~al.}(2022){Dotti}, {Bonetti}, {Rigamonti}, {Bortolas}, {Fossati}, {Decarli}, {Covino}, {Lupi}, {Franchini}, {Sesana}, \& {Calderone}}]{2022arXiv220506275D}
{Dotti}, M., {Bonetti}, M., {Rigamonti}, F., {et~al.} 2022, arXiv e-prints, arXiv:2205.06275.
\newblock \doarXiv{2205.06275}

\bibitem[{{Drake} {et~al.}(2009){Drake}, {Djorgovski}, {Mahabal}, {Beshore}, {Larson}, {Graham}, {Williams}, {Christensen}, {Catelan}, {Boattini}, {Gibbs}, {Hill}, \& {Kowalski}}]{2009ApJ...696..870D}
{Drake}, A.~J., {Djorgovski}, S.~G., {Mahabal}, A., {et~al.} 2009, \apj, 696, 870, \dodoi{10.1088/0004-637X/696/1/870}

\bibitem[{{Duras} {et~al.}(2020){Duras}, {Bongiorno}, {Ricci}, {Piconcelli}, {Shankar}, {Lusso}, {Bianchi}, {Fiore}, {Maiolino}, {Marconi}, {Onori}, {Sani}, {Schneider}, {Vignali}, \& {La Franca}}]{fred}
{Duras}, F., {Bongiorno}, A., {Ricci}, F., {et~al.} 2020, \aap, 636, A73, \dodoi{10.1051/0004-6361/201936817}

\bibitem[{{Eracleous} {et~al.}(2012){Eracleous}, {Boroson}, {Halpern}, \& {Liu}}]{2012ApJS..201...23E}
{Eracleous}, M., {Boroson}, T.~A., {Halpern}, J.~P., \& {Liu}, J. 2012, \apjs, 201, 23, \dodoi{10.1088/0067-0049/201/2/23}

\bibitem[{{Farris} {et~al.}(2014){Farris}, {Duffell}, {MacFadyen}, \& {Haiman}}]{2014ApJ...783..134F}
{Farris}, B.~D., {Duffell}, P., {MacFadyen}, A.~I., \& {Haiman}, Z. 2014, \apj, 783, 134, \dodoi{10.1088/0004-637X/783/2/134}

\bibitem[{{Farris} {et~al.}(2015){Farris}, {Duffell}, {MacFadyen}, \& {Haiman}}]{2015MNRAS.447L..80F}
---. 2015, \mnras, 447, L80, \dodoi{10.1093/mnrasl/slu184}

\bibitem[{{Fruscione} {et~al.}(2006){Fruscione}, {McDowell}, {Allen}, {Brickhouse}, {Burke}, {Davis}, {Durham}, {Elvis}, {Galle}, {Harris}, {Huenemoerder}, {Houck}, {Ishibashi}, {Karovska}, {Nicastro}, {Noble}, {Nowak}, {Primini}, {Siemiginowska}, {Smith}, \& {Wise}}]{2006SPIE.6270E..1VF}
{Fruscione}, A., {McDowell}, J.~C., {Allen}, G.~E., {et~al.} 2006, in Society of Photo-Optical Instrumentation Engineers (SPIE) Conference Series, Vol. 6270, Society of Photo-Optical Instrumentation Engineers (SPIE) Conference Series, ed. D.~R. {Silva} \& R.~E. {Doxsey}, 62701V, \dodoi{10.1117/12.671760}

\bibitem[{{Gabriel} {et~al.}(2004){Gabriel}, {Denby}, {Fyfe}, {Hoar}, {Ibarra}, {Ojero}, {Osborne}, {Saxton}, {Lammers}, \& {Vacanti}}]{2004ASPC..314..759G}
{Gabriel}, C., {Denby}, M., {Fyfe}, D.~J., {et~al.} 2004, in Astronomical Society of the Pacific Conference Series, Vol. 314, Astronomical Data Analysis Software and Systems (ADASS) XIII, ed. F.~{Ochsenbein}, M.~G. {Allen}, \& D.~{Egret}, 759

\bibitem[{{Gehrels} {et~al.}(2004){Gehrels}, {Chincarini}, {Giommi}, {Mason}, {Nousek}, {Wells}, {White}, {Barthelmy}, {Burrows}, {Cominsky}, {Hurley}, {Marshall}, {M{\'e}sz{\'a}ros}, {Roming}, {Angelini}, {Barbier}, {Belloni}, {Campana}, {Caraveo}, {Chester}, {Citterio}, {Cline}, {Cropper}, {Cummings}, {Dean}, {Feigelson}, {Fenimore}, {Frail}, {Fruchter}, {Garmire}, {Gendreau}, {Ghisellini}, {Greiner}, {Hill}, {Hunsberger}, {Krimm}, {Kulkarni}, {Kumar}, {Lebrun}, {Lloyd-Ronning}, {Markwardt}, {Mattson}, {Mushotzky}, {Norris}, {Osborne}, {Paczynski}, {Palmer}, {Park}, {Parsons}, {Paul}, {Rees}, {Reynolds}, {Rhoads}, {Sasseen}, {Schaefer}, {Short}, {Smale}, {Smith}, {Stella}, {Tagliaferri}, {Takahashi}, {Tashiro}, {Townsley}, {Tueller}, {Turner}, {Vietri}, {Voges}, {Ward}, {Willingale}, {Zerbi}, \& {Zhang}}]{2004ApJ...611.1005G}
{Gehrels}, N., {Chincarini}, G., {Giommi}, P., {et~al.} 2004, \apj, 611, 1005, \dodoi{10.1086/422091}

\bibitem[{{Gierli{\'n}ski} \& {Done}(2004)}]{2004MNRAS.349L...7G}
{Gierli{\'n}ski}, M., \& {Done}, C. 2004, \mnras, 349, L7, \dodoi{10.1111/j.1365-2966.2004.07687.x}

\bibitem[{{Gorenstein} \& {Tucker}(1976)}]{1976ARA&A..14..373G}
{Gorenstein}, P., \& {Tucker}, W.~H. 1976, \araa, 14, 373, \dodoi{10.1146/annurev.aa.14.090176.002105}

\bibitem[{{Graham} {et~al.}(2015{\natexlab{a}}){Graham}, {Djorgovski}, {Stern}, {Glikman}, {Drake}, {Mahabal}, {Donalek}, {Larson}, \& {Christensen}}]{2015Natur.518...74G}
{Graham}, M.~J., {Djorgovski}, S.~G., {Stern}, D., {et~al.} 2015{\natexlab{a}}, \nat, 518, 74, \dodoi{10.1038/nature14143}

\bibitem[{{Graham} {et~al.}(2015{\natexlab{b}}){Graham}, {Djorgovski}, {Stern}, {Drake}, {Mahabal}, {Donalek}, {Glikman}, {Larson}, \& {Christensen}}]{2015MNRAS.453.1562G}
---. 2015{\natexlab{b}}, \mnras, 453, 1562, \dodoi{10.1093/mnras/stv1726}

\bibitem[{{Gualandris} {et~al.}(2017){Gualandris}, {Read}, {Dehnen}, \& {Bortolas}}]{2017MNRAS.464.2301G}
{Gualandris}, A., {Read}, J.~I., {Dehnen}, W., \& {Bortolas}, E. 2017, \mnras, 464, 2301, \dodoi{10.1093/mnras/stw2528}

\bibitem[{{Guo} {et~al.}(2019){Guo}, {Liu}, {Shen}, {Loeb}, {Monroe}, \& {Prochaska}}]{2019MNRAS.482.3288G}
{Guo}, H., {Liu}, X., {Shen}, Y., {et~al.} 2019, \mnras, 482, 3288, \dodoi{10.1093/mnras/sty2920}

\bibitem[{{Guti{\'e}rrez} {et~al.}(2022){Guti{\'e}rrez}, {Combi}, {Noble}, {Campanelli}, {Krolik}, {L{\'o}pez Armengol}, \& {Garc{\'\i}a}}]{2022ApJ...928..137G}
{Guti{\'e}rrez}, E.~M., {Combi}, L., {Noble}, S.~C., {et~al.} 2022, \apj, 928, 137, \dodoi{10.3847/1538-4357/ac56de}

\bibitem[{{Harrison} {et~al.}(2013){Harrison}, {Craig}, {Christensen}, {Hailey}, {Zhang}, {Boggs}, {Stern}, {Cook}, {Forster}, {Giommi}, {Grefenstette}, {Kim}, {Kitaguchi}, {Koglin}, {Madsen}, {Mao}, {Miyasaka}, {Mori}, {Perri}, {Pivovaroff}, {Puccetti}, {Rana}, {Westergaard}, {Willis}, {Zoglauer}, {An}, {Bachetti}, {Barri{\`e}re}, {Bellm}, {Bhalerao}, {Brejnholt}, {Fuerst}, {Liebe}, {Markwardt}, {Nynka}, {Vogel}, {Walton}, {Wik}, {Alexander}, {Cominsky}, {Hornschemeier}, {Hornstrup}, {Kaspi}, {Madejski}, {Matt}, {Molendi}, {Smith}, {Tomsick}, {Ajello}, {Ballantyne}, {Balokovi{\'c}}, {Barret}, {Bauer}, {Blandford}, {Brandt}, {Brenneman}, {Chiang}, {Chakrabarty}, {Chenevez}, {Comastri}, {Dufour}, {Elvis}, {Fabian}, {Farrah}, {Fryer}, {Gotthelf}, {Grindlay}, {Helfand}, {Krivonos}, {Meier}, {Miller}, {Natalucci}, {Ogle}, {Ofek}, {Ptak}, {Reynolds}, {Rigby}, {Tagliaferri}, {Thorsett}, {Treister}, \& {Urry}}]{2013ApJ...770..103H}
{Harrison}, F.~A., {Craig}, W.~W., {Christensen}, F.~E., {et~al.} 2013, \apj, 770, 103, \dodoi{10.1088/0004-637X/770/2/103}

\bibitem[{{Jansen} {et~al.}(2001){Jansen}, {Lumb}, {Altieri}, {Clavel}, {Ehle}, {Erd}, {Gabriel}, {Guainazzi}, {Gondoin}, {Much}, {Munoz}, {Santos}, {Schartel}, {Texier}, \& {Vacanti}}]{2001A&A...365L...1J}
{Jansen}, F., {Lumb}, D., {Altieri}, B., {et~al.} 2001, \aap, 365, L1, \dodoi{10.1051/0004-6361:20000036}

\bibitem[{{Ju} {et~al.}(2013){Ju}, {Greene}, {Rafikov}, {Bickerton}, \& {Badenes}}]{2013ApJ...777...44J}
{Ju}, W., {Greene}, J.~E., {Rafikov}, R.~R., {Bickerton}, S.~J., \& {Badenes}, C. 2013, \apj, 777, 44, \dodoi{10.1088/0004-637X/777/1/44}

\bibitem[{{Jun} {et~al.}(2015){Jun}, {Stern}, {Graham}, {Djorgovski}, {Mainzer}, {Cutri}, {Drake}, \& {Mahabal}}]{2015ApJ...814L..12J}
{Jun}, H.~D., {Stern}, D., {Graham}, M.~J., {et~al.} 2015, \apjl, 814, L12, \dodoi{10.1088/2041-8205/814/1/L12}

\bibitem[{{Kamraj} {et~al.}(2022){Kamraj}, {Brightman}, {Harrison}, {Stern}, {Garc{\'\i}a}, {Balokovi{\'c}}, {Ricci}, {Koss}, {Mej{\'\i}a-Restrepo}, {Oh}, {Powell}, \& {Urry}}]{2022ApJ...927...42K}
{Kamraj}, N., {Brightman}, M., {Harrison}, F.~A., {et~al.} 2022, \apj, 927, 42, \dodoi{10.3847/1538-4357/ac45f6}

\bibitem[{{Komossa} {et~al.}(2020){Komossa}, {Grupe}, {Parker}, {Valtonen}, {G{\'o}mez}, {Gopakumar}, \& {Dey}}]{2020MNRAS.498L..35K}
{Komossa}, S., {Grupe}, D., {Parker}, M.~L., {et~al.} 2020, \mnras, 498, L35, \dodoi{10.1093/mnrasl/slaa125}

\bibitem[{{Komossa} {et~al.}(2023){Komossa}, {Grupe}, {Kraus}, {Gurwell}, {Haiman}, {Liu}, {Tchekhovskoy}, {Gallo}, {Berton}, {Blandford}, {G{\'o}mez}, \& {Gonzalez}}]{2023MNRAS.522L..84K}
{Komossa}, S., {Grupe}, D., {Kraus}, A., {et~al.} 2023, \mnras, 522, L84, \dodoi{10.1093/mnrasl/slad016}

\bibitem[{{Koss} {et~al.}(2017){Koss}, {Trakhtenbrot}, {Ricci}, {Lamperti}, {Oh}, {Berney}, {Schawinski}, {Balokovi{\'c}}, {Baronchelli}, {Crenshaw}, {Fischer}, {Gehrels}, {Harrison}, {Hashimoto}, {Hogg}, {Ichikawa}, {Masetti}, {Mushotzky}, {Sartori}, {Stern}, {Treister}, {Ueda}, {Veilleux}, \& {Winter}}]{2017ApJ...850...74K}
{Koss}, M., {Trakhtenbrot}, B., {Ricci}, C., {et~al.} 2017, \apj, 850, 74, \dodoi{10.3847/1538-4357/aa8ec9}

\bibitem[{{Krause} {et~al.}(2019){Krause}, {Shabala}, {Hardcastle}, {Bicknell}, {B{\"o}hringer}, {Chon}, {Nawaz}, {Sarzi}, \& {Wagner}}]{2019MNRAS.482..240K}
{Krause}, M. G.~H., {Shabala}, S.~S., {Hardcastle}, M.~J., {et~al.} 2019, \mnras, 482, 240, \dodoi{10.1093/mnras/sty2558}

\bibitem[{{Krolik} {et~al.}(2019){Krolik}, {Volonteri}, {Dubois}, \& {Devriendt}}]{2019ApJ...879..110K}
{Krolik}, J.~H., {Volonteri}, M., {Dubois}, Y., \& {Devriendt}, J. 2019, \apj, 879, 110, \dodoi{10.3847/1538-4357/ab24c9}

\bibitem[{{Kun} {et~al.}(2014){Kun}, {Gab{\'a}nyi}, {Karouzos}, {Britzen}, \& {Gergely}}]{2014MNRAS.445.1370K}
{Kun}, E., {Gab{\'a}nyi}, K.~{\'E}., {Karouzos}, M., {Britzen}, S., \& {Gergely}, L.~{\'A}. 2014, \mnras, 445, 1370, \dodoi{10.1093/mnras/stu1813}

\bibitem[{{Lehto} \& {Valtonen}(1996)}]{1996ApJ...460..207L}
{Lehto}, H.~J., \& {Valtonen}, M.~J. 1996, \apj, 460, 207, \dodoi{10.1086/176962}

\bibitem[{{Li} {et~al.}(2016){Li}, {Wang}, {Ho}, {Lu}, {Qiu}, {Du}, {Hu}, {Huang}, {Zhang}, {Wang}, \& {Bai}}]{2016ApJ...822....4L}
{Li}, Y.-R., {Wang}, J.-M., {Ho}, L.~C., {et~al.} 2016, \apj, 822, 4, \dodoi{10.3847/0004-637X/822/1/4}

\bibitem[{{Liao} {et~al.}(2021){Liao}, {Chen}, {Liu}, {Holgado}, {Guo}, {Gruendl}, {Morganson}, {Shen}, {Davis}, {Kessler}, {Martini}, {McMahon}, {Allam}, {Annis}, {Avila}, {Banerji}, {Bechtol}, {Bertin}, {Brooks}, {Buckley-Geer}, {Carnero Rosell}, {Carrasco Kind}, {Carretero}, {Javier Castander}, {Cunha}, {D'Andrea}, {da Costa}, {Davis}, {De Vicente}, {Desai}, {Thomas Diehl}, {Doel}, {Eifler}, {Evrard}, {Flaugher}, {Fosalba}, {Frieman}, {Garcia-Bellido}, {Gaztanaga}, {Glazebrook}, {Gruen}, {Gschwend}, {Gutierrez}, {Hartley}, {Hollowood}, {Honscheid}, {Hoyle}, {James}, {Krause}, {Kuehn}, {Lima}, {Maia}, {Marshall}, {Menanteau}, {Miquel}, {Plazas Malag{\'o}n}, {Roodman}, {Sanchez}, {Scarpine}, {Schubnell}, {Serrano}, {Smith}, {Smith}, {Soares-Santos}, {Sobreira}, {Suchyta}, {Swanson}, {Tarle}, {Vikram}, \& {Walker}}]{2021MNRAS.500.4025L}
{Liao}, W.-T., {Chen}, Y.-C., {Liu}, X., {et~al.} 2021, \mnras, 500, 4025, \dodoi{10.1093/mnras/staa3055}

\bibitem[{{Liu} {et~al.}(2018){Liu}, {Gezari}, \& {Miller}}]{2018ApJ...859L..12L}
{Liu}, T., {Gezari}, S., \& {Miller}, M.~C. 2018, \apjl, 859, L12, \dodoi{10.3847/2041-8213/aac2ed}

\bibitem[{{Liu} {et~al.}(2019){Liu}, {Gezari}, {Ayers}, {Burgett}, {Chambers}, {Hodapp}, {Huber}, {Kudritzki}, {Metcalfe}, {Tonry}, {Wainscoat}, \& {Waters}}]{2019ApJ...884...36L}
{Liu}, T., {Gezari}, S., {Ayers}, M., {et~al.} 2019, \apj, 884, 36, \dodoi{10.3847/1538-4357/ab40cb}

\bibitem[{{Liu} {et~al.}(2020){Liu}, {Koss}, {Blecha}, {Ricci}, {Trakhtenbrot}, {Mushotzky}, {Harrison}, {Ichikawa}, {Kakkad}, {Oh}, {Powell}, {Privon}, {Schawinski}, {Shimizu}, {Smith}, {Stern}, {Treister}, \& {Urry}}]{2020ApJ...896..122L}
{Liu}, T., {Koss}, M., {Blecha}, L., {et~al.} 2020, \apj, 896, 122, \dodoi{10.3847/1538-4357/ab952d}

\bibitem[{{Liu} {et~al.}(2014){Liu}, {Shen}, {Bian}, {Loeb}, \& {Tremaine}}]{2014ApJ...789..140L}
{Liu}, X., {Shen}, Y., {Bian}, F., {Loeb}, A., \& {Tremaine}, S. 2014, \apj, 789, 140, \dodoi{10.1088/0004-637X/789/2/140}

\bibitem[{{Lobanov} \& {Roland}(2005)}]{2005A&A...431..831L}
{Lobanov}, A.~P., \& {Roland}, J. 2005, \aap, 431, 831, \dodoi{10.1051/0004-6361:20041831}

\bibitem[{{Lusso} \& {Risaliti}(2016)}]{2016ApJ...819..154L}
{Lusso}, E., \& {Risaliti}, G. 2016, \apj, 819, 154, \dodoi{10.3847/0004-637X/819/2/154}

\bibitem[{{Lusso} {et~al.}(2010){Lusso}, {Comastri}, {Vignali}, {Zamorani}, {Brusa}, {Gilli}, {Iwasawa}, {Salvato}, {Civano}, {Elvis}, {Merloni}, {Bongiorno}, {Trump}, {Koekemoer}, {Schinnerer}, {Le Floc'h}, {Cappelluti}, {Jahnke}, {Sargent}, {Silverman}, {Mainieri}, {Fiore}, {Bolzonella}, {Le F{\`e}vre}, {Garilli}, {Iovino}, {Kneib}, {Lamareille}, {Lilly}, {Mignoli}, {Scodeggio}, \& {Vergani}}]{2010A&A...512A..34L}
{Lusso}, E., {Comastri}, A., {Vignali}, C., {et~al.} 2010, \aap, 512, A34, \dodoi{10.1051/0004-6361/200913298}

\bibitem[{{Madsen} {et~al.}(2015){Madsen}, {Harrison}, {Markwardt}, {An}, {Grefenstette}, {Bachetti}, {Miyasaka}, {Kitaguchi}, {Bhalerao}, {Boggs}, {Christensen}, {Craig}, {Forster}, {Fuerst}, {Hailey}, {Perri}, {Puccetti}, {Rana}, {Stern}, {Walton}, {J{\o}rgen Westergaard}, \& {Zhang}}]{2015ApJS..220....8M}
{Madsen}, K.~K., {Harrison}, F.~A., {Markwardt}, C.~B., {et~al.} 2015, \apjs, 220, 8, \dodoi{10.1088/0067-0049/220/1/8}

\bibitem[{{Magdziarz} \& {Zdziarski}(1995)}]{1995MNRAS.273..837M}
{Magdziarz}, P., \& {Zdziarski}, A.~A. 1995, \mnras, 273, 837, \dodoi{10.1093/mnras/273.3.837}

\bibitem[{{Mayer} {et~al.}(2007){Mayer}, {Kazantzidis}, {Madau}, {Colpi}, {Quinn}, \& {Wadsley}}]{2007Sci...316.1874M}
{Mayer}, L., {Kazantzidis}, S., {Madau}, P., {et~al.} 2007, Science, 316, 1874, \dodoi{10.1126/science.1141858}

\bibitem[{{McKernan} {et~al.}(2013){McKernan}, {Ford}, {Kocsis}, \& {Haiman}}]{2013MNRAS.432.1468M}
{McKernan}, B., {Ford}, K.~E.~S., {Kocsis}, B., \& {Haiman}, Z. 2013, \mnras, 432, 1468, \dodoi{10.1093/mnras/stt567}

\bibitem[{{Mehdipour} {et~al.}(2011){Mehdipour}, {Branduardi-Raymont}, {Kaastra}, {Petrucci}, {Kriss}, {Ponti}, {Blustin}, {Paltani}, {Cappi}, {Detmers}, \& {Steenbrugge}}]{2011A&A...534A..39M}
{Mehdipour}, M., {Branduardi-Raymont}, G., {Kaastra}, J.~S., {et~al.} 2011, \aap, 534, A39, \dodoi{10.1051/0004-6361/201116875}

\bibitem[{{Nandra} \& {Pounds}(1994)}]{1994MNRAS.268..405N}
{Nandra}, K., \& {Pounds}, K.~A. 1994, \mnras, 268, 405, \dodoi{10.1093/mnras/268.2.405}

\bibitem[{{Nasa High Energy Astrophysics Science Archive Research Center (Heasarc)}(2014)}]{2014ascl.soft08004N}
{Nasa High Energy Astrophysics Science Archive Research Center (Heasarc)}. 2014, {HEAsoft: Unified Release of FTOOLS and XANADU}, Astrophysics Source Code Library, record ascl:1408.004.
\newblock \doeprint{1408.004}

\bibitem[{{O'Neill} {et~al.}(2022){O'Neill}, {Kiehlmann}, {Readhead}, {Aller}, {Blandford}, {Liodakis}, {Lister}, {Mr{\'o}z}, {O'Dea}, {Pearson}, {Ravi}, {Vallisneri}, {Cleary}, {Graham}, {Grainge}, {Hodges}, {Hovatta}, {L{\"a}hteenm{\"a}ki}, {Lamb}, {Lazio}, {Max-Moerbeck}, {Pavlidou}, {Prince}, {Reeves}, {Tornikoski}, {Vergara de la Parra}, \& {Zensus}}]{2022ApJ...926L..35O}
{O'Neill}, S., {Kiehlmann}, S., {Readhead}, A.~C.~S., {et~al.} 2022, \apjl, 926, L35, \dodoi{10.3847/2041-8213/ac504b}

\bibitem[{{Pandey} \& {Singh}(2008)}]{2008MNRAS.387.1627P}
{Pandey}, J.~C., \& {Singh}, K.~P. 2008, \mnras, 387, 1627, \dodoi{10.1111/j.1365-2966.2008.13342.x}

\bibitem[{{Qian} {et~al.}(2018){Qian}, {Britzen}, {Witzel}, {Krichbaum}, \& {Kun}}]{2018A&A...615A.123Q}
{Qian}, S.~J., {Britzen}, S., {Witzel}, A., {Krichbaum}, T.~P., \& {Kun}, E. 2018, \aap, 615, A123, \dodoi{10.1051/0004-6361/201732039}

\bibitem[{{Ricci} {et~al.}(2017){Ricci}, {Trakhtenbrot}, {Koss}, {Ueda}, {Del Vecchio}, {Treister}, {Schawinski}, {Paltani}, {Oh}, {Lamperti}, {Berney}, {Gandhi}, {Ichikawa}, {Bauer}, {Ho}, {Asmus}, {Beckmann}, {Soldi}, {Balokovi{\'c}}, {Gehrels}, \& {Markwardt}}]{2017ApJS..233...17R}
{Ricci}, C., {Trakhtenbrot}, B., {Koss}, M.~J., {et~al.} 2017, \apjs, 233, 17, \dodoi{10.3847/1538-4365/aa96ad}

\bibitem[{{Robrade} \& {Schmitt}(2005)}]{2005A&A...435.1073R}
{Robrade}, J., \& {Schmitt}, J.~H.~M.~M. 2005, \aap, 435, 1073, \dodoi{10.1051/0004-6361:20041941}

\bibitem[{{Roedig} {et~al.}(2014){Roedig}, {Krolik}, \& {Miller}}]{2014ApJ...785..115R}
{Roedig}, C., {Krolik}, J.~H., \& {Miller}, M.~C. 2014, \apj, 785, 115, \dodoi{10.1088/0004-637X/785/2/115}

\bibitem[{{Ryan} \& {MacFadyen}(2017)}]{2017ApJ...835..199R}
{Ryan}, G., \& {MacFadyen}, A. 2017, \apj, 835, 199, \dodoi{10.3847/1538-4357/835/2/199}

\bibitem[{{Saade} {et~al.}(2020){Saade}, {Stern}, {Brightman}, {Haiman}, {Djorgovski}, {D'Orazio}, {Ford}, {Graham}, {Jun}, {Kraft}, {McKernan}, {Vikhlinin}, \& {Walton}}]{2020ApJ...900..148S}
{Saade}, M.~L., {Stern}, D., {Brightman}, M., {et~al.} 2020, \apj, 900, 148, \dodoi{10.3847/1538-4357/abad31}

\bibitem[{{Serafinelli} {et~al.}(2020){Serafinelli}, {Severgnini}, {Braito}, {Della Ceca}, {Vignali}, {Ambrosino}, {Cicone}, {Zaino}, {Dotti}, {Sesana}, {Gianolli}, {Ballo}, {La Parola}, \& {Matzeu}}]{2020ApJ...902...10S}
{Serafinelli}, R., {Severgnini}, P., {Braito}, V., {et~al.} 2020, \apj, 902, 10, \dodoi{10.3847/1538-4357/abb3c3}

\bibitem[{{Sesana} {et~al.}(2018){Sesana}, {Haiman}, {Kocsis}, \& {Kelley}}]{2018ApJ...856...42S}
{Sesana}, A., {Haiman}, Z., {Kocsis}, B., \& {Kelley}, L.~Z. 2018, \apj, 856, 42, \dodoi{10.3847/1538-4357/aaad0f}

\bibitem[{{Shemmer} {et~al.}(2008){Shemmer}, {Brandt}, {Netzer}, {Maiolino}, \& {Kaspi}}]{2008ApJ...682...81S}
{Shemmer}, O., {Brandt}, W.~N., {Netzer}, H., {Maiolino}, R., \& {Kaspi}, S. 2008, \apj, 682, 81, \dodoi{10.1086/588776}

\bibitem[{{Shen} {et~al.}(2013){Shen}, {Liu}, {Loeb}, \& {Tremaine}}]{2013ApJ...775...49S}
{Shen}, Y., {Liu}, X., {Loeb}, A., \& {Tremaine}, S. 2013, \apj, 775, 49, \dodoi{10.1088/0004-637X/775/1/49}

\bibitem[{{Tang} {et~al.}(2018){Tang}, {Haiman}, \& {MacFadyen}}]{2018MNRAS.476.2249T}
{Tang}, Y., {Haiman}, Z., \& {MacFadyen}, A. 2018, \mnras, 476, 2249, \dodoi{10.1093/mnras/sty423}

\bibitem[{{Tsai} {et~al.}(2013){Tsai}, {Jarrett}, {Stern}, {Emonts}, {Barrows}, {Assef}, {Norris}, {Eisenhardt}, {Lonsdale}, {Blain}, {Benford}, {Wu}, {Stalder}, {Stubbs}, {High}, {Li}, \& {Kong}}]{2013ApJ...779...41T}
{Tsai}, C.-W., {Jarrett}, T.~H., {Stern}, D., {et~al.} 2013, \apj, 779, 41, \dodoi{10.1088/0004-637X/779/1/41}

\bibitem[{{Valtonen} {et~al.}(2008){Valtonen}, {Lehto}, {Nilsson}, {Heidt}, {Takalo}, {Sillanp{\"a}{\"a}}, {Villforth}, {Kidger}, {Poyner}, {Pursimo}, {Zola}, {Wu}, {Zhou}, {Sadakane}, {Drozdz}, {Koziel}, {Marchev}, {Ogloza}, {Porowski}, {Siwak}, {Stachowski}, {Winiarski}, {Hentunen}, {Nissinen}, {Liakos}, \& {Dogru}}]{2008Natur.452..851V}
{Valtonen}, M.~J., {Lehto}, H.~J., {Nilsson}, K., {et~al.} 2008, \nat, 452, 851, \dodoi{10.1038/nature06896}

\bibitem[{{Vaughan} {et~al.}(2016){Vaughan}, {Uttley}, {Markowitz}, {Huppenkothen}, {Middleton}, {Alston}, {Scargle}, \& {Farr}}]{2016MNRAS.461.3145V}
{Vaughan}, S., {Uttley}, P., {Markowitz}, A.~G., {et~al.} 2016, \mnras, 461, 3145, \dodoi{10.1093/mnras/stw1412}

\bibitem[{{Walton} {et~al.}(2013){Walton}, {Nardini}, {Fabian}, {Gallo}, \& {Reis}}]{2013MNRAS.428.2901W}
{Walton}, D.~J., {Nardini}, E., {Fabian}, A.~C., {Gallo}, L.~C., \& {Reis}, R.~C. 2013, \mnras, 428, 2901, \dodoi{10.1093/mnras/sts227}

\bibitem[{{Weisskopf} {et~al.}(2002){Weisskopf}, {Brinkman}, {Canizares}, {Garmire}, {Murray}, \& {Van Speybroeck}}]{2002PASP..114....1W}
{Weisskopf}, M.~C., {Brinkman}, B., {Canizares}, C., {et~al.} 2002, \pasp, 114, 1, \dodoi{10.1086/338108}

\bibitem[{{Xin} {et~al.}(2020){Xin}, {Charisi}, {Haiman}, {Schiminovich}, {Graham}, {Stern}, \& {D'Orazio}}]{2020MNRAS.496.1683X}
{Xin}, C., {Charisi}, M., {Haiman}, Z., {et~al.} 2020, \mnras, 496, 1683, \dodoi{10.1093/mnras/staa1643}

\bibitem[{{Xin} {et~al.}(2021){Xin}, {Mingarelli}, \& {Hazboun}}]{2021ApJ...915...97X}
{Xin}, C., {Mingarelli}, C. M.~F., \& {Hazboun}, J.~S. 2021, \apj, 915, 97, \dodoi{10.3847/1538-4357/ac01c5}

\bibitem[{{XRISM Science Team}(2020)}]{2020arXiv200304962X}
{XRISM Science Team}. 2020, arXiv e-prints, arXiv:2003.04962, \dodoi{10.48550/arXiv.2003.04962}

\bibitem[{{Zhu} \& {Thrane}(2020)}]{2020ApJ...900..117Z}
{Zhu}, X.-J., \& {Thrane}, E. 2020, \apj, 900, 117, \dodoi{10.3847/1538-4357/abac5a}

\end{thebibliography}
\nocite{*}
\bibliographystyle{aasjournal}

\end{document}